\documentclass[10pt,journal,compsoc]{IEEEtran}
\usepackage[utf8x]{inputenc}
\usepackage[T1]{fontenc}
\usepackage[scaled]{inconsolata}
\usepackage[american]{babel}
\usepackage{cite}
\usepackage{colortbl}
\usepackage{booktabs}
\usepackage{tabularx}
\usepackage{graphicx}
\usepackage{xspace}
\usepackage{makecell}
\usepackage{mathtools}
\usepackage{multirow}
\usepackage[normalem]{ulem}
\usepackage{adjustbox}
\usepackage[spacing,stretch=10,shrink=40]{microtype}
\usepackage[acronyms,nohypertypes={acronym},nomain]{glossaries}
\usepackage[binary-units, detect-all, per-mode=symbol, range-units=single, range-phrase=~to~]{siunitx}
\usepackage[dvipsnames,table]{xcolor}
\usepackage{amssymb}
\usepackage{amsbsy}
\usepackage{tikz}
\ifCLASSOPTIONcompsoc
  \usepackage[caption=false,font=normalsize,labelfont=sf,textfont=sf]{subfig}
\else
  \usepackage[caption=false,font=footnotesize]{subfig}
\fi

\usepackage[hidelinks]{hyperref}
\usepackage{cleveref}

\newif\ifthesis

\newif\ifrevhl
\ifrevhl
  \colorlet{colorrevhl}{Blue}
  
\else
  \colorlet{colorrevhl}{black}
  
\fi

\DeclarePairedDelimiter{\floor}{\lfloor}{\rfloor}

\DeclareSIUnit{\dpflop}{dpflop}
\DeclareSIUnit{\GE}{GE}
\DeclareSIUnit{\x}{\!$\times$}

\crefname{section}{\S}{\S\S}
\crefname{figure}{Fig.\@}{Figs.\@}
\crefname{table}{Table}{Tables}

\newacronym{abi}{ABI}{application binary interface}
\newacronym{api}{API}{application programming interface}
\newacronym{asic}{ASIC}{application-specific integrated circuit}\glsunset{asic}
\newacronym{axi}{AXI}{Advanced eXtensible Interface}
\newacronym{clb}{CLB}{configurable logic block}
\newacronym{cnn}{CNN}{convolutional neural network}
\newacronym{cpu}{CPU}{central processing unit}\glsunset{cpu}
\newacronym{csr}{CSR}{control and status register}
\newacronym{ddr}{DDR}{double data rate}\glsunset{ddr}
\newacronym{dma}{DMA}{direct memory access}
\newacronym{dram}{DRAM}{dynamic random access memory}\glsunset{dram}
\newacronym{dwc}{DWC}{data width converter}
\newacronym{eda}{EDA}{electronic design automation}
\newacronym{elf}{ELF}{executable and linkable format}
\newacronym{fame}{FAME}{FPGA Architecture Model Execution}
\newacronym{ff}{FF}{flip-flop}
\newacronym{fifo}{FIFO}{first-in first-out buffer}
\newacronym{flop}{FLOP}{floating-point operation}
\newacronym{fmc}{FMC}{FPGA Mezzanine Card}
\newacronym{fpga}{FPGA}{field-programmable gate array}\glsunset{fpga}
\newacronym{fpu}{FPU}{floating-point unit}
\newacronym{ge}{GE}{gate equivalent}
\newacronym{gpgpu}{GPGPU}{general-purpose graphics processing unit}\glsunset{gpgpu}
\newacronym{gpu}{GPU}{graphics processing unit}\glsunset{gpu}
\newacronym{hal}{HAL}{hardware abstraction library}
\newacronym{hbm}{HBM}{high bandwidth memory}
\newacronym{io}{I/O}{input/output}
\newacronym{iommu}{IOMMU}{I/O memory management unit}\glsunset{iommu}
\newacronym[longplural={intellectual properties}]{ip}{IP}{intellectual property}
\newacronym{ir}{IR}{intermediate representation}
\newacronym{isa}{ISA}{instruction set architecture}
\newacronym{iss}{ISS}{instruction set simulation}
\newacronym{lsu}{LSU}{load/store unit}
\newacronym{lto}{LTO}{link-time optimization}
\newacronym{mac}{MAC}{multiply-accumulate}
\newacronym{mlt}{MLT}{machine learning training}
\newacronym{mpsoc}{MPSoC}{multiprocessor system-on-chip}
\newacronym{msb}{MSB}{most significant bit}
\newacronym{mmu}{MMU}{memory management unit}
\newacronym{nic}{NIC}{network interface controller}
\newacronym{nn}{NN}{neural network}
\newacronym[longplural={networks-on-chip}]{noc}{NoC}{network-on-chip}
\newacronym{os}{OS}{operating system}
\newacronym{pl}{PL}{programmable logic}
\newacronym{pmca}{PMCA}{programmable many-core accelerator}
\newacronym{rtl}{RTL}{register-transfer level}
\newacronym{simd}{SIMD}{single instruction multiple data}
\newacronym[longplural={systems-on-chip}]{soc}{SoC}{system-on-chip}
\newacronym{spm}{SPM}{scratch-pad memory}
\newacronym{sram}{SRAM}{static random access memory}\glsunset{sram}
\newacronym{smmu}{SMMU}{system memory management unit}
\newacronym{tcdm}{TCDM}{tightly-coupled data memory}
\newacronym{tlb}{TLB}{translation lookaside buffer}
\newacronym{vmm}{VMM}{virtual memory management}

\newcommand{\darknet}{\texttt{darknet}\xspace}
\newcommand{\hero}{HEROv2\xspace}
\newcommand{\herovone}{HEROv1\xspace}
\newcommand{\host}{host\xspace}
\newcommand{\Host}{Host\xspace}

\newcommand{\bigO}[1]{\ensuremath{\mathcal{O}\left(#1\right)}}
\newcommand{\mineq}{\mathrel{-}=}

\begin{document}

\bstctlcite{IEEEexample:BSTcontrol}

\setlength{\floatsep}{.6\baselineskip plus .2\baselineskip minus .2\baselineskip}
\setlength{\textfloatsep}{.7\baselineskip plus .2\baselineskip minus .3 \baselineskip}
\setlength{\dblfloatsep}{\floatsep}
\setlength{\dbltextfloatsep}{\textfloatsep}
\setlength{\abovecaptionskip}{.25\baselineskip}

\newcommand{\titleraw}{HEROv2: Full-Stack Open-Source Research Platform for Heterogeneous Computing}
\newcommand{\titlefmt}{\titleraw}
\title{\titlefmt}

\ifx\anonymous\undefined
\author{%
  Andreas~Kurth,~\IEEEmembership{Student Member,~IEEE,}
  Bj\"{o}rn~Forsberg, %
  and~Luca~Benini,~\IEEEmembership{Fellow,~IEEE}
}
\else
\author{\textit{Anonymous Author(s)}}
\fi
\markboth{%
}{%
  \ifx\anonymous\undefined%
    Kurth \MakeLowercase{\textit{et al.}}: \titleraw%
  \fi%
}

\IEEEtitleabstractindextext{%
\begin{abstract}
Heterogeneous computers integrate general-purpose host processors with domain-specific accelerators to combine versatility with efficiency and high performance.
To realize the full potential of heterogeneous computers, however, many hardware and software design challenges have to be overcome.
While architectural and system simulators can be used to analyze heterogeneous computers, they are faced with unavoidable compromises between simulation speed and performance modeling accuracy.

In this work we present \hero, an \acrshort{fpga}-based research platform that enables accurate and fast exploration of heterogeneous computers consisting of accelerators based on clusters of 32-bit RISC-V cores and an application-class 64-bit ARMv8 or RV64 host processor.
\hero allows to seamlessly share data between 64-bit \host{}s and 32-bit accelerators and comes with a fully open-source on-chip network, a unified heterogeneous programming interface, and a mixed-data-model, mixed-\acrshort{isa} heterogeneous compiler based on LLVM.
We evaluate \hero in four case studies from the application level over toolchain and system architecture down to accelerator microarchitecture.
We demonstrate how \hero enables effective research and development on the full stack of heterogeneous computing.
For instance, the compiler can tile loops and infer data transfers to and from the accelerators, which leads to a speedup of up to \SI{4.4}{\x} compared to the original program and in most cases is only \SI{15}{\percent} slower than a handwritten implementation, which requires \SI{2.6}{\x} more code.
\end{abstract}
}

\maketitle

\IEEEdisplaynontitleabstractindextext

\IEEEpeerreviewmaketitle

\ifthesis\else\IEEEraisesectionheading{\section{Introduction}\label{sec:introduction}}\fi

\noindent
\IEEEPARstart{H}{eterogeneous} integrated computing systems aim to combine general-purpose computing with domain-specific, efficient processing capabilities~\cite{horowitz2014energyproblem,zahran2017heterogeneous,dally2020dsa}.
Such computers integrate a general-purpose \host processor with specialized programmable many-core accelerators (e.g., \cite{nvidia2018xavier,tesla2019fsd,amd2020renoir}).
These systems are very complex and many challenges remain to be overcome to realize their full potential~\cite{hennessy2019goldenage}.
Central questions over the full stack of computing, from application programming over compilers and runtime libraries down to the accelerator microarchitecture, include:
How to partition tasks between host and different accelerators?
How to express that partitioning in programming languages, optimize it in the toolchain, or both?
How to manage data sharing across host and accelerator, share address spaces and overcome the differences between cache-coherent memory subsystems, typically found on the \host, and their non-coherent counterparts, which are typically found in accelerators?
Which types and combinations of accelerators are optimal for a given domain?

Research on heterogeneous systems traditionally follows a double-track approach, where accelerators are developed in isolation~\cite{reuther2020mlaccels,gui2019graphaccels}, and their impact on system-level performance is estimated through analytical models and simulators~\cite{ubal2012multi2sim,power2015gem5gpu}.
Compared to using a prototype heterogeneous system, this approach has significant drawbacks:
First, interactions between host, accelerators, the memory hierarchy, and peripherals are complex to model accurately, making accurate simulation orders of magnitude slower than running prototypes.
Second, even full-system simulators model heterogeneous computers to a limited degree only~\cite{butko2016simulation}.
For example, models of system-level interconnects or \glspl{smmu}, are missing or highly abstract and imprecise.
Third, simulators that are not precisely calibrated and accuracy-validated against the simulated system are generally too inaccurate to provide reliable results, and full-system simulators are particularly unreliable~\cite{akram2019simulation}.
A research platform that serves as a working prototype, on the other hand, enables collaborative and accurate architectural analysis and optimization~\cite{lee2016agile}.
To perform system-level research using standard benchmarks and real-world applications, the platform must additionally provide a software stack that includes an application programming interface and a complete compiler toolchain.

Existing research platforms do not meet these requirements in their entirety.
Many provide a custom accelerator on programmable logic~\cite{gray2016grviphalanx,kamaleldin2020riscvfpga}, and some even couple the accelerator to a host processor that runs an operating system~\cite{mantovani2020openesp,balkind2020openpiton}.
\herovone~\cite{kurth2018herov1} provides software stack and compiler that enable the evaluation of real-world applications on a mixed-\gls{isa} computer, but it fundamentally restricts host and accelerator to use the same data model (e.g., 32-bit).
Additionally, \herovone's on-chip network and memory subsystem are restricted to simple architectures that cannot meet the demands of modern heterogeneous computers.

In this work, we make three main contributions:
\begin{enumerate}
\item
We resolve the mentioned limitations and present \hero{}, an open-source research platform where an application-class 64-bit host can seamlessly share data with a 32-bit parallel programmable accelerator.
The latter is implemented on programmable logic and based on permissively licensed open-source \acrshort{rtl}\footnote{%
    \acrshort{rtl} = \acrlong{rtl}.
    We use the industry-standard SystemVerilog hardware description language.
} components.
The hardware components can be freely extended and modified and include a high-performance end-to-end on-chip network that can be fully customized to meet the memory demands of the accelerator and target application~(\cref{sec:platform:hw_arch}).
The platform also includes a complete heterogeneous compiler based on LLVM, which allows single-source single-binary development of heterogeneous applications with OpenMP~4.5 offloading~(\cref{sec:platform:toolchain}).
The runtime libraries on the accelerator and driver on the host enable this offloading with little overhead~(\cref{sec:platform:runtime}).
A unified heterogeneous \gls{api} enables productive programming while providing fine-grained control where necessary~(\cref{sec:platform:api}).
The complete software stack and tools are open source under a permissive license.
\item
We demonstrate the capabilities of \hero{} by using it to study four
current research topics in heterogeneous computing
and provide quantitative insights on the level of applications~(\cref{sec:eval:app}), toolchains~(\cref{sec:eval:toolchain}), system architecture~(\cref{sec:eval:system}), and accelerator architecture~(\cref{sec:eval:accel}).
\item
We also leverage \hero{} to design and evaluate a novel solution to one of the most pressing problems in heterogeneous computing: how to relieve the programmer of the burden of specializing an algorithm to the memory hierarchy of the accelerator~(\cref{sec:eval:toolchain}).
\end{enumerate}
Section 2 describes \hero{}'s hardware and software platform.
Section 3 focuses on case studies.
Furthermore, we compare with related work in \cref{sec:related_work} and conclude in \cref{sec:conclusion}.

\section{Platform}%
\label{sec:platform}

\noindent
\hero{} consists of a complete heterogeneous hardware architecture~(\cref{sec:platform:hw_arch}) as well as an end-to-end software stack including a toolchain and compilers~(\cref{sec:platform:toolchain}), operating system and runtime libraries~(\cref{sec:platform:runtime}), and an \acrlong{api}~(\cref{sec:platform:api}).

\subsection{Hardware Architecture}%
\label{sec:platform:hw_arch}

\begin{figure*}[tbp]
  \centering
  \includegraphics[width=\ifthesis\else.95\fi\textwidth]{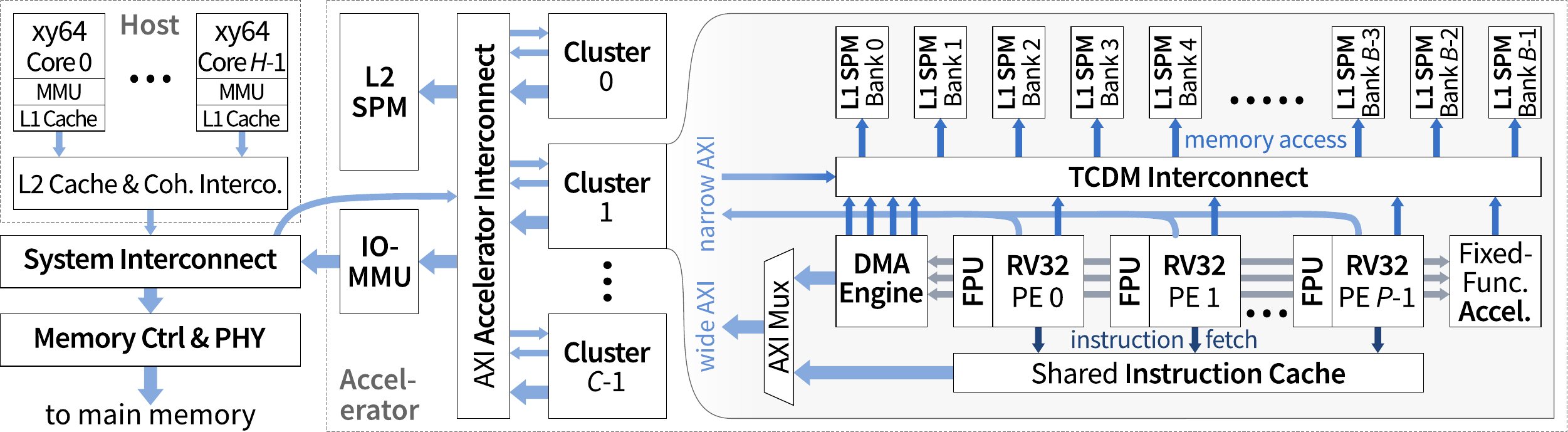}
  \caption{%
    \hero{}'s hardware architecture, which combines a general-purpose \emph{host} processor (in the upper left corner) with a domain-specific programmable many-core \emph{accelerator} (on the right side) so that data in main memory can be shared effectively (in the lower left corner).
  }%
  \label{fig:platform:hw_arch}
\end{figure*}

\noindent
\hero{}'s hardware architecture combines a general-purpose \gls{cpu} running a full \gls{os} with a domain-specific programmable multi- or many-core accelerator.
\Cref{fig:platform:hw_arch} gives an overview of the components and their connections.
As many hardware components as possible are implemented on an \gls{fpga} (also called \emph{programmable logic}) to make them configurable, modifiable, and extendable.
All hardware modules mapped on the \gls{fpga} are available in synthesizable \gls{rtl} logic under a permissive open-source license, which makes them fully analyzable and freely extensible and reusable.
The vast majority of hardware components is silicon-proven, meaning they have been and will be used in \gls{asic} tapeouts in many modern silicon technologies.

The host \gls{cpu} is a hard-macro 64-bit ARMv8 Cortex-A multi-core on Xilinx' Zynq UltraScale+ family \glspl{fpga} or a soft-macro 64-bit RISC-V core (CVA6 architecture~\cite{zaruba2019ariane}) on any UltraScale+ family \gls{fpga}.
The general design principle of the host core is to maximize performance per area or power on mostly sequential workloads with a complex control flow.
Each host core features a private L1 instruction cache, L1 data cache, and a \gls{mmu}.
All host cores are attached to a coherent interconnect and share an L2 data and instruction cache.

Host and accelerator share an off-chip main memory (\gls{ddr} \gls{dram} or \gls{hbm}) through the system interconnect, which can be coherent to the caches of the host.
The memory hierarchy of the accelerator consists of software-managed \glspl{spm}.
To copy data into and out of the \glspl{spm}, the accelerator features \gls{dma} engines.
To share the virtual address space of an application running on the host, each accelerator features a hybrid \gls{iommu} (such as \cite{vogel2017onaccelptw}).
This \gls{iommu} consists mainly of a \gls{tlb}, which translates virtual user-space application addresses to physical memory addresses and supports a high degree of concurrency (e.g., tens of outstanding transactions from different masters).
The \gls{tlb} is managed by the accelerator itself, which handles \gls{tlb} misses by walking the application page table managed by the host and filling the corresponding entries into the \gls{tlb}.
The \gls{iommu} is called \emph{hybrid} because it is managed in software, which allows the accelerator to efficiently share virtual address pointers with a minimum amount of hardware (e.g., for buffers).

The accelerator is composed of many minimal 32-bit RISC-V cores, which are organized into clusters of 4 to 16 cores for scalability.
Different RISC-V core architectures are supported (see \cref{tbl:eval:platforms}), and consequentially the specific \gls{isa} of the accelerator varies, but all accelerator cores support at least the \texttt{RV32IMA} \gls{isa}.
The focus of the accelerator core architecture is to maximize the performance per area or power on computation-heavy workloads with a simple control flow.
For this reason, the cores feature a single-issue in-order pipeline with 1 to 4 stages.
To accelerate floating-point workloads, each core can be extended with a \gls{fpu}, which is highly parametrizable:
depending on the needs of the application, it can execute one double-precision (fp64) \gls{mac}, one or two single-precision (fp32) \glspl{mac}, two to four half-precision (fp16) \glspl{mac}, or four to eight quarter-precision (fp8) \glspl{mac} in one clock cycle~\cite{mach2021fpnew}.

To accelerate workloads that heavily rely on functions outside common integer or floating-point operations, the cores support custom bit-manipulation instructions, and the cluster can additionally be extended with fixed-function hardware processing engines (e.g., \cite{conti2018xne}).
To maximize the utilization of the compute units, each accelerator core supports custom instructions to repeat a sequence of instructions multiple times without branches (so-called \emph{hardware loops}) as well as custom instructions to implicitly increment the memory address on a load or store.

Within each accelerator cluster, the cores have single-cycle access to a multi-banked, tightly-coupled L1 data \gls{spm}.
A default banking factor of two allows any core to access any bank in any cycle with a low probability of contention for most applications.
The cores can additionally access memory outside the own cluster, including shared main memory, with a latency between a few (to other clusters) to hundreds of cycles (to main memory, depending on on-chip network and memory controller).
A custom \gls{csr} allows each 32-bit core to load from and store to any 64-bit address~\cite[\S~5]{kurth2020cc}.
This \gls{csr} extends the native 32-bit address by 32 upper \si{\bit} and is set automatically by the compiler (see \cref{sec:platform:host_accel_interop}).

The cores fetch their instructions from an L1 instruction cache, which is shared by all cores in one cluster.
To reduce the pressure on the shared instruction cache during loops, each core additionally contains an L0 instruction cache holding up to eight compressed instructions.

Finally, each accelerator cluster features a \gls{dma} engine, which can address the full 64-bit memory space, supports unified virtual memory through the hybrid \gls{iommu}~\cite{kurth2018iccd}, can transfer up to \SI{1024}{\bit} per clock cycle in and out of the cluster (full duplex), and can have tens of transactions, each consisting of tens of data beats, outstanding at any time.
This \gls{dma} engine allows to transfer data in high-bandwidth bursts while the accelerator cores compute on data in local memory.
If an application allows issuing sufficiently many or long bursts, the \gls{dma} engine allows tolerating a latency of hundreds of cycles between main memory and accelerator, which is crucial to support the ongoing trend of deeper and non-uniform memory hierarchies.

Multiple accelerator clusters are interconnected with two non-coherent networks: a wide one for high-bandwidth \gls{dma} transfers and a narrow one for low-latency accesses by cores~\cite{kurth2021axi}.
A high-bandwidth on-chip \gls{sram} controller connects the L2 \gls{spm}, which is shared by all clusters, to the accelerator interconnect.
The hybrid \gls{iommu} connects the accelerator to the host.
The \gls{iommu} is either directly attached to a non-coherent system interconnect or via a bridge (such as \cite{cavalcante2020bridge}) as I/O-coherent request node to a coherent system interconnect.
To program and control the accelerator from the host, it is additionally connected as I/O-coherent slave node to the system interconnect.

\subsection{Toolchain and Compilers}%
\label{sec:platform:toolchain}

\begin{figure*}[tbp]
    \centering
    \includegraphics[width=\ifthesis\else0.9\fi\textwidth]{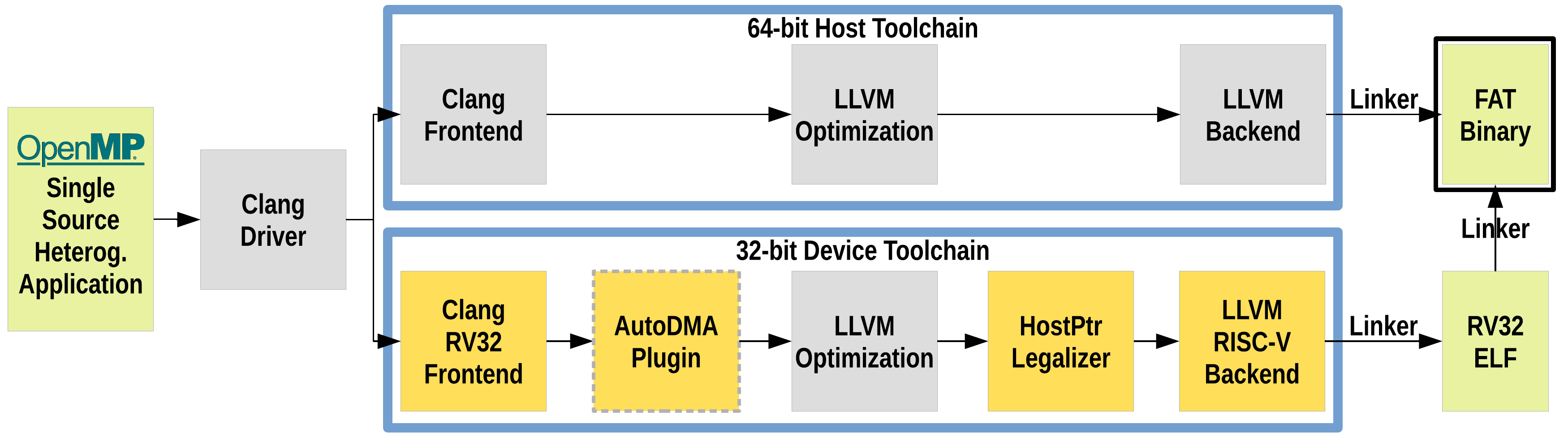}
    \caption{Overview of HEROv2's heterogeneous toolchain and compilers.}
    \label{fig:platform:toolchain}
\end{figure*}

\hero{}'s heterogeneous hardware requires toolchain support to enable the development of applications for the platform in an efficient and productive way.
\hero{} provides such a heterogeneous toolchain, based on LLVM~9\footnote{%
  An update to LLVM 12 is planned in the near future.
}, which provides efficient support for heterogeneous compilation based on OpenMP.
This enables the seamless co-integration of compute-focused accelerator kernels and control-focused host code into a unified application, including target-specific compilation and optimizations.
Additionally, the different data widths of the system (64-bit host and 32-bit accelerator) are supported by LLVM's address space implementation, which provides the compiler with the means to express pointers of varying width.
An overview of \hero{}'s toolchain is shown in \cref{fig:platform:toolchain}.

The toolchain flow starts by compiling OpenMP-annotated heterogeneous source code, as shown at the left of the figure.
OpenMP describes heterogeneity and parallelism through \texttt{\#pragma}s and leaves the transformation to parallel code to the compiler.
Annotating a piece of code with \texttt{\#pragma omp target} directs the toolchain to compile the code both for the host\footnote{%
  By OpenMP's specification, the runtime decides during execution time if a target region is executed on the host or the accelerator, but in the case of \hero{}, the latter is always the case.
} and the accelerator.
We refer to these regions as \emph{target regions}.
The respective host and device toolchains are thereby invoked by the \emph{Clang driver}, as shown in \cref{fig:platform:toolchain}, which transform the source code into an object file for each architecture.
The Clang driver triggers the device linker for the device object file, creating a RISC-V \gls{elf} file, whereafter the host linker is triggered.
The host linker first links the host object files into an \gls{elf} file and then embeds the device \gls{elf} as an object inside the host \gls{elf}, creating a \emph{FAT binary}.
This allows the OpenMP runtime to load the device \gls{elf} into accelerator memory at runtime.

\hero{} uses an off-the-shelf LLVM-based 64-bit toolchain for the host and a custom LLVM-based 32-bit RISC-V toolchain for the device.
Both toolchains are marked in \cref{fig:platform:toolchain}, where the components that include customizations are highlighted in yellow tones, and aim to provide interoperability between host and accelerators, ease of programming, and support for \gls{isa} extensions.

\subsubsection{Interoperability between Host and Accelerators}
\label{sec:platform:host_accel_interop}

Pointers in C/C++, as well as in the LLVM \gls{ir}, have a fixed width: the \emph{data width} of the target processor.
A 32-bit data width of an accelerator therefore implies that 64-bit pointers from the host will be truncated.
To allow 64-bit host pointers to be correctly represented, an additional 64-bit \emph{address space} is defined in \hero{}'s accelerator compiler.
We refer to the two address spaces as the 32-bit \emph{native address space} and the 64-bit \emph{host address space}.
Address space support is a built-in LLVM feature, and has been previously used, e.g., to separate pointers to global and shared memory in CUDA.
In such cases, however, pointers are annotated by the programmer, e.g., \texttt{\_\_shared\_\_} in CUDA, and all pointers typically have the same width.

To address mixed-data-width compilation in \hero{}, the \emph{Clang frontend} has been extended to generate LLVM \gls{ir} with automatically assigned address spaces.
We adopt the techniques of \cite{kurth2020cc}, where OpenMP offloading entry points are used to infer that pointers passed to a device kernel from the host are 64-bit wide.
The use of such pointers are then tracked throughout the application, such that any pointer that \emph{cannot} be guaranteed to never hold a 64-bit host address is \emph{promoted} to the host address space.
Any pointer that is guaranteed to only hold 32-bit pointers is kept in the native address space.
Additional control is handed to the programmer through \texttt{\_\_device} pointer decorations, to enforce a pointer to belong to the native address space, if the compiler could not guarantee it to be correct.
As part of machine code generation, any data types and operations that are not natively supported by the underlying hardware and/or \gls{abi} must be \emph{legalized}.
As pointer semantics are dropped in LLVM backends (i.e., pointers are treated as integers), the backend is able to implicitly legalize arithmetic operations on 64-bit pointers.
However, the backend does not support the legalization of wider-than-native load and store operations.
\hero{}'s RISC-V compiler has therefore been extended with a custom \emph{host pointer legalizer} pass right before the optimized code is passed to the RISC-V backend for machine code generation.
This pass identifies all load and store operations on addresses in the host address space and implements them using the address extension \gls{csr}.

\subsubsection{Ease of Programming and Code Portability}

An important aspect for code portability and ease of pro\-gram\-ming is the automatic optimization of code for the memory hierarchy of a computer.
\hero{}'s accelerators use software-managed \glspl{spm}, which are refilled using \gls{dma} engines.
This means software must explicitly orchestrate any data movements between shared main memory and fast local memory.
As OpenMP does not provide any mechanisms to tile data structures and move tiles with \gls{dma} transfers, programmers need to manually rewrite their code to perform well on \gls{spm}-based accelerators.
\hero{}'s \gls{dma} \gls{api} is unified over all accelerators, but the initial tiling of an application is nonetheless a significant effort and reduces code portability outside \hero{}.

To reduce this effort and improve code portability, \hero{}'s device compiler provides an optional \emph{AutoDMA} plugin that automatically analyzes source code to identify memory regions that are suitable for staging through \glspl{spm} and transforms the code to automatically program the \gls{dma} engine without any programmer intervention.
The AutoDMA plugin is also able to perform loop tiling to extract segments of code whose memory footprint is small enough to fit in the local memory.
The AutoDMA plugin is an extension of \emph{HePREM}~\cite{forsberg2020heprem}, originally envisioned for transforming real-time \gls{gpu} code to be less sensitive to memory interference.
This was achieved by transforming \gls{gpu} kernels into a series of \emph{load}, \emph{execute}, and \emph{store} phases, with explicit synchronization points between them.
These three phases are well aligned with accelerators based on software-managed \glspl{spm}.
In contrast to \emph{HePREM}, which targets \gls{dma}-less \gls{gpu} systems, \emph{AutoDMA} generates \gls{dma} \gls{api} calls instead of moving data using load and store instructions.
Additionally, synchronization has been minimized to improve performance.
The resulting \emph{AutoDMA} plugin provides an optional way to achieve performance on \hero{} without the need for manual tiling and \gls{dma} management code.

\subsubsection{Support for \texorpdfstring{\acrshort{isa}}{ISA} Extensions}

The device compiler backend of \hero{} has been extended to support the \gls{isa} extensions supported by the RV32 cores of the accelerator.
This includes the automatic detection and insertion of \emph{hardware loop} instructions, automatic optimization to generate \emph{post-increment} load and store instructions, as well as pattern matching to emit \emph{multiply-accumulate} instructions, outlined in \cref{sec:platform:hw_arch}.
To the best of our knowledge, this is the first time custom instructions have been implemented for RISC-V in LLVM, and a full-system performance evaluation is shown in the case study in \cref{sec:eval:accel}.

\vspace*{\baselineskip}
In summary, \hero{}'s heterogeneous toolchain provides a de-facto standard and heterogeneous-by-design programming model via OpenMP, which is fully support by its LLVM-based compiler.
This provides seamless, end-to-end single-source-to-heterogeneous-binary compilation.
The device compiler has been significantly extended to support performance, ease of programming, and code portability:
first, through the minimization of expensive wider-than-native load and store operations in a mixed data-model setting;
second, through the support for automatic tiling and \gls{dma} management through \emph{AutoDMA};
and third, through automatic code generation targeting the performance-oriented \gls{isa} extensions supported by the underlying hardware.

\subsection{Runtime Libraries and Operating System Support}%
\label{sec:platform:runtime}

\begin{figure}[htbp]
  \centering
  \includegraphics[width=\ifthesis.7\else.85\fi\columnwidth]{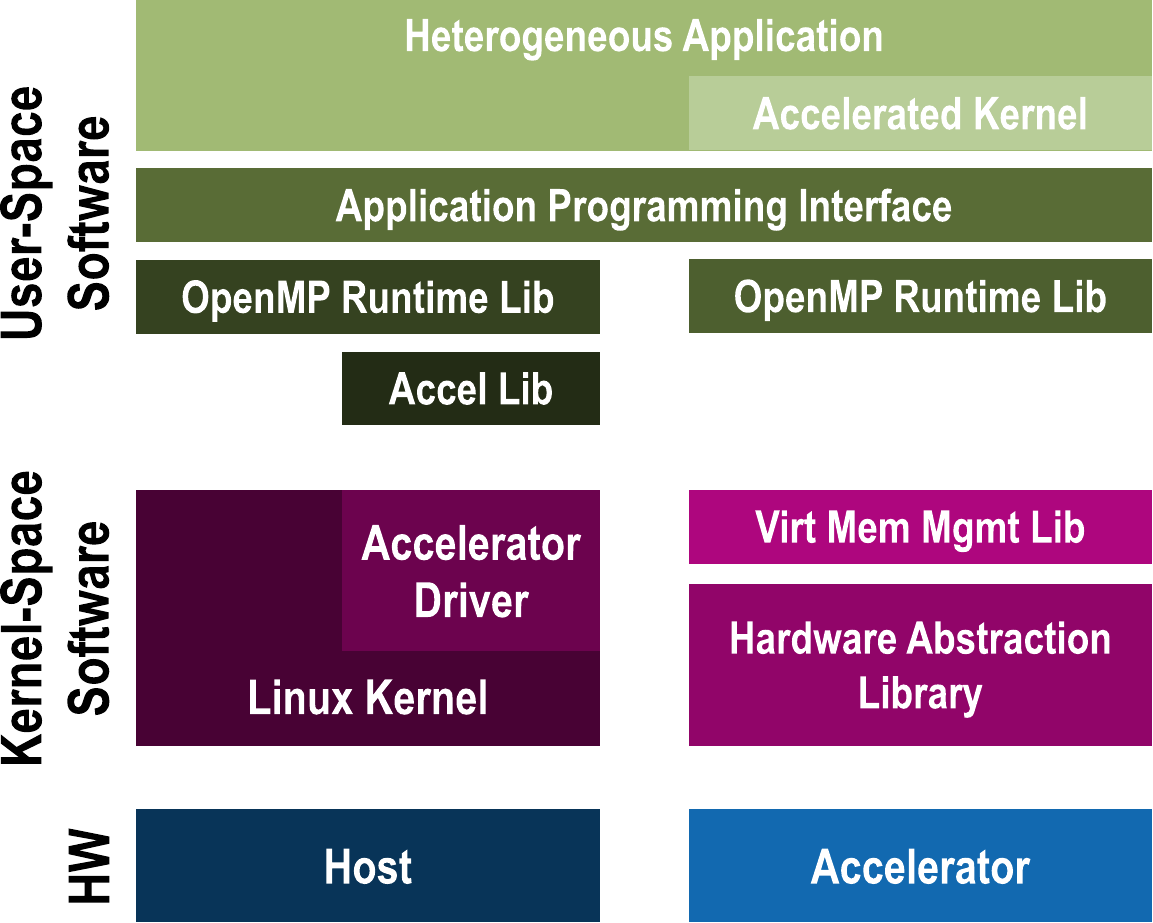}
  \caption{%
    \hero{}'s runtime stack, which seamlessly integrates \emph{accelerators} (with their runtime stack on the right) into the \acrshort{os} running on the \emph{host} (runtime stack on the left) to enable heterogeneous applications with transparent offloading of accelerated kernels (at the top).
  }%
  \label{fig:platform:runtime}
\end{figure}

\noindent
\hero{}'s runtime software stack is designed to seamlessly integrate the accelerators into the \gls{os} running on the host and allow for transparent accelerator programming with OpenMP~4.5 offloading~\cite{openmp45} and unified virtual memory compliant with HSA specifications~\cite{hsa2016}.
An overview of the runtime stack is shown in \cref{fig:platform:runtime}.
This section discusses the layers below the \gls{api}, which is discussed separately in \cref{sec:platform:api}.

A heterogeneous application starts executing on the host.
When the host encounters a \texttt{\#pragma omp target} directive, it offloads the code within the \texttt{target} region to the specified (or default) accelerator (called \emph{device} in OpenMP terminology).
To this end, the host OpenMP runtime library invokes the device-specific runtime plugin.
The plugin passes a pointer to the offloaded code and data to a hardware mailbox in the device, thereby starting execution on the device.
The first core of the first cluster of the device runs an \emph{offload manager}.
It is woken by an interrupt from the hardware mailbox and starts executing the offloaded function.
All data items inside the \texttt{map} clause become available to the device:
When unified virtual memory is enabled, pointers are passed unmodified, no data is copied, and the device is given read-only access to the user-space page table of the application on the host.
Otherwise, the host copies data to a physically contiguous memory region in main memory and changes the pointers before passing them to the device.
By design, offloading does \emph{not} copy data to the \glspl{spm} of the device.
There are two main reasons for this:
First, \hero{}'s accelerator model aims at accelerating kernels that take at least a few ten thousand cycles to execute.
Thus, the offloading model is relatively coarse grained and the \texttt{map}ped data in its entirety in general does not fit into the local memory of the device.
Second, OpenMP's \texttt{map} clauses cannot express tiling, yet flexible tiling is essential for efficient execution on device-local memory.

Inside the offloaded region, execution starts on the first core of the first cluster.
When that core encounters a \texttt{\#pragma omp teams} directive, it forks execution to multiple clusters, and the cluster master core (i.e., the first core of each cluster) starts executing the region.
When a cluster master core encounters a \texttt{\#pragma omp parallel} directive, it forks execution to multiple cores of its cluster.
Inside parallel regions, all OpenMP worksharing, datasharing, and synchronization constructs are available, allowing for effective parallel programming following OpenMP's standard paradigm.
The OpenMP device runtime library implements the \texttt{\_\_kmpc\_*} functions emitted by the compiler by calling into the accelerator-specific \gls{hal}.

The \gls{vmm} library allows the accelerator to share the virtual address space of a user-space application running on the host (concept of \cite{vogel2017onaccelptw}).
After the host has set up entries in the \gls{iommu} that allow the accelerator to access the page table, the \gls{vmm} library provides functions to translate any valid virtual address to a physical address and set up a corresponding translation entry in the hybrid \gls{iommu}.
Application programmers usually do not notice this:
The compiler generates the correct instructions for accessing pointers outside the native (32-bit physical) address space of the accelerator.
In the common case, such accesses hit in the \gls{tlb} of the \gls{iommu} and incur an overhead of only three cycles per remote memory access~\cite{kurth2020cc}.
When an access misses in the \gls{tlb}, the core either invokes the \gls{vmm} library itself to add an entry to the \gls{iommu}, or it lets a dedicated core handle the misses.
The latter is preferable for pointer-based applications, and miss handling can be configured per offload through custom options to the \texttt{target} region.
The implementation of the \gls{vmm} library is specific to the virtual memory system of the host (e.g., ARM VMSAv8-64 or RISC-V Sv39 or Sv48).

The \gls{hal} on the accelerator provides functions for forking parallel execution, identifying and synchronizing cores, putting cores to sleep and waking them up, controlling the \gls{dma} engine, and communicating between clusters and with the host through the mailbox.
The \gls{hal} is implemented using low-level hardware-specific primitives, such as writing memory-mapped registers and setting bits in \glspl{csr}.

The \gls{os} device driver and the accompanying user-space accelerator library on the host implement the accelerator-specific functionality for offloading to and communicating with the accelerator from the host.
This includes identifying the accelerator in the device tree, resetting, initializing, and programming it, and making the page table of the user-space process readable for the accelerator.

In summary, \hero{}'s modular runtime stack supports different hosts and accelerators while reusing large parts of the code base, combining flexibility with accelerator-specific specialization.
On the accelerator, all runtime libraries are linked into the offloaded application, and \gls{lto} minimizes the overhead of the multiple layers.
On the host, system calls are required to trigger and conclude an offload, but the overhead of that is negligible due to \hero{}'s coarse-grained offloading model.

\subsection{Application Programming Interface}%
\label{sec:platform:api}

\noindent
The application level is the most important from the perspective of end users and application developers.
\hero{}'s toolchain and runtime software provide the means to make effective use of accelerators, but without a properly designed \gls{api} applications on heterogeneous computers remain too complex to program in most cases.
Porting an application to make efficient use of the software-managed memory of an accelerator involves tiling data and scheduling data transfers, which are difficult tasks in general.
An \gls{api} alone cannot solve this problem, but it can make the work of the application programmer portable over different accelerators and substantially easier by abstracting the intricacies of the hardware away.
The design goal is to provide an interface that is unified over all supported accelerators together with an implementation that is optimized and verified for each accelerator individually.
\hero{}'s \gls{api} complements the OpenMP \gls{api} for offloading and parallel programming~(\cref{sec:platform:runtime}) and the accelerator-specific compiler~(\cref{sec:platform:toolchain}), which optimizes the compute part of an application for the target accelerator.

\hero{}'s \gls{api} has three main categories of functionality: memory management for the different \gls{spm} levels, data transfers between \glspl{spm} and main memory, and performance measurements.
All functions are thread-safe and can thus be used inside and outside parallel regions.

To manage the heap memory of the accelerator, there are three functions for each \gls{spm} level:
\texttt{hero\_lN\_capacity} returns the currently available heap memory at \gls{spm} level $N$.
This function is often used at the beginning of a tiling region to calculate the tile sizes.
\texttt{hero\_lN\_malloc} and \texttt{hero\_lN\_free} implement POSIX' memory allocation and freeing functions~\cite{posix2017} for \gls{spm} level $N$.
The implementation uses a deterministic constant-complexity memory allocator~\cite{herter2014allocation,kirienko2020o1heap}, ensures mutual exclusivity among all affected cores (e.g., within the same cluster for L1 \gls{spm}) through RISC-V atomic operations, and can detect heap overflows with a canary mechanism.
The alignment and minimum allocation granule is \SI{8}{\byte}.

To transfer data between \glspl{spm} and main memory, \hero{} provides multiple functions with the semantics of POSIX' \texttt{memcpy}~\cite{posix2017}.
Those functions are organized in three dimensions: direction (device-to-host or host-to-device), synchronicity (blocking or asynchronous), and transfer dimensionality (1D, 2D, etc.).
The direction has to be distinguished in the function signature because pointers and addresses in the host-managed main memory are of a different width and address space than device-internal pointers: in \texttt{hero\_memcpy\_host2dev\_*} functions, the \texttt{src} pointer is in the host address space and the \texttt{dst} pointer in the device address space, and vice-versa for the \texttt{hero\_memcpy\_dev2host\_*} functions.

The synchronicity distinguishes functions that return as soon as the \gls{dma} engine has been programmed (with \texttt{\_async} suffix) or after all data has been transferred (without suffix).
The asynchronous functions allow to start a \gls{dma} transfer and then work on different data while the \gls{dma} engine completes the transfer.
Those functions return a unique transfer identifier, which has to be passed to the \texttt{hero\_memcpy\_wait} function to guarantee transfer completion before the data can be used.
Multi-dimensional transfers allow to scatter and gather non-contiguous data with a single function call.
For instance, the \texttt{hero\_memcpy2d\_*} functions copy $N$ sequences of $B$ bytes from \texttt{src} to \texttt{dst} and apply a different address offset to \texttt{src} and \texttt{dst} after each sequence.
This scatter-gather functionality is essential for tiling (e.g., to gather the rows of a tile of a 2D matrix from main memory into a dense \gls{spm} buffer before computation and scatter them back after computation).
Whenever the \gls{dma} engine supports multi-dimensional transfers, they are executed directly by the \gls{dma} hardware; otherwise, they are implemented in software.

To measure the performance of applications and their execution on hardware, \hero{} provides functions that provide a uniform interface to different hardware performance monitors and counters.
The functions are mainly designed for hardware counters to which an event is assigned dynamically, which is common in modern processors.
The available events range from monotonic clock cycles over memory accesses and stalls to memory and interconnect contention and utilization metrics.
The \texttt{hero\_perf\_alloc} function allocates a counter for a given event and resets that counter.
If the event is not supported by the hardware or the hardware counters are exhausted, the function returns an error.
At the start of a program section to be investigated, a call to \texttt{hero\_perf\_continue\_all} starts all allocated counters, and at the end of that section, \texttt{hero\_perf\_pause\_all} stops them.
Those two functions execute with the minimal latency and overhead supported by the hardware (often as a single inlined \gls{csr} write instruction), allowing for precise, fine-grained, and minimally intrusive performance measurements, which are crucial for identifying bottlenecks and systematic optimization.

\section{Evaluation}%
\label{sec:eval}

\begin{table}[htbp]
  \centering
  \maxsizebox{\columnwidth}{!}{%
  \begin{tabular}{ l | c | c | c }
    \toprule
    \textbf{Configuration} & \textbf{Aurora} & \textbf{Blizzard} & \textbf{Cyclone} \\
    \midrule
    \textbf{Host ISA} & \multicolumn{2}{c|}{ARMv8.0-A} & \texttt{RV64GC} \\
    \textbf{Host Core Arch.} & \multicolumn{2}{c|}{Cortex-A53} & CVA6~\cite{zaruba2019ariane} \\
    \textbf{Host \# Cores} & \multicolumn{2}{c|}{4} & 1 \\
    \textbf{Accel.\ ISA} & \texttt{RV32IMAFCXpulpv2} & \multicolumn{2}{c}{\texttt{RV32IMAFDXssrXfrepXsdma}} \\
    \textbf{Accel.\ Core Arch} & CV32E40P~\cite{gautschi2017core} & \multicolumn{2}{c}{Snitch~\cite{zaruba2020snitch}} \\
    \textbf{Accel.\ \# Cores} & \multicolumn{2}{c|}{8} & 32 \\
    \textbf{Main Mem.\ Cap.} & \multicolumn{2}{c|}{\SI{4}{\gibi\byte} DDR4} & \SI{8}{\gibi\byte} HBM2E \\
    \textbf{Main Mem. BW} & \multicolumn{2}{c|}{up to \SI{19.2}{\giga\byte\per\second}} & up to \SI{460}{\giga\byte\per\second} \\
    \textbf{Carrier Silicon} & \multicolumn{2}{c|}{Xilinx ZU9EG} & Xilinx VU37P \\
    \textbf{Carrier Freq.} & \multicolumn{2}{c|}{\SI{50}{\mega\hertz}} & \SI{25}{\mega\hertz} \\
    \textbf{Status} & mature & \multicolumn{2}{c}{in development} \\
    \bottomrule
  \end{tabular}
  }
  \caption{Current target platforms and configurations of \hero{}.}%
  \label{tbl:eval:platforms}
\end{table}

\noindent
In this section, we evaluate the most mature configuration of \hero:
As \host, it features an industry-standard quad-core 64-bit ARMv8 Cortex-A53 processor with \SI{32}{\kibi\byte} L1 instruction and \SI{32}{\kibi\byte} L1 data cache per core and an \SI{1}{\mebi\byte} L2 cache shared by all four cores, implemented as hard macro and clocked at \SI{1.2}{\giga\hertz}.
As \gls{pmca}, it features an octa-core 32-bit RISC-V floating-point accelerator (OpenHW CV32E40P core architecture) with \SI{128}{\kibi\byte} L1 \gls{spm} and support for custom instructions (\texttt{RV32IMAFCXpulpv2}), implemented as soft-macro in the \gls{pl}.
\Host and \gls{pmca} are connected through a lightweight \gls{iommu}, which allows the \gls{pmca} to share the \host's virtual memory space and which is implemented as soft-macro in \gls{pl}, to a shared \gls{dram} controller.
The shared main memory consists of \SI{4}{\gibi\byte} DDR4 \gls{dram}, which provides up to \SI{19.2}{\giga\byte\per\second} of bandwidth.

The implementation of \gls{pmca} and \gls{iommu} on the \gls{pl} of a Xilinx Zynq UltraScale+ ZU9EG \gls{soc} achieves a clock frequency of \SI{50}{\mega\hertz} (without any \gls{fpga}-specific optimizations).
The frequency is mainly limited by paths from the \textit{request} output of the \gls{lsu} of an accelerator core through the cluster interconnect to the arbitrator of a memory bank and back to \textit{grant} input of the \gls{lsu} of another core.
Among the available \gls{pl} resources, the \glspl{clb} are the limiting factor with \SI{98.1}{\percent} utilization, of which \SI{87.7}{\percent} are used by the \gls{pmca} and \SI{10.4}{\percent} by the \gls{iommu}.
Within the \gls{pmca}, the cores (each of which includes an \gls{fpu}), dominate with \SI{38.4}{\percent} of the total \glspl{clb}.
\SI{24.2}{\percent} of the block RAM tiles and \SI{2.9}{\percent} of the DSP slices are used.
We used Xilinx Vivado~2019.2 with the \textit{Alternate Routability} synthesis strategy and the \textit{Congestion--Spread Logic--Low} implementation strategy.

Variants of \hero with alternative \host processors and \glspl{pmca} are in development, and an overview of current configurations of \hero and their status is shown in \cref{tbl:eval:platforms}.
The \emph{Blizzard} configuration shares the host and the carrier silicon with the \emph{Aurora} configuration evaluated here but features an octa-core RISC-V \gls{mlt} accelerator (\texttt{RV32IMAFDXssrXfrepXsdma}) with variable precision support for \SIrange{8}{64}{\bit} floating-point numbers.
The \emph{Cyclone} configuration targets a larger carrier silicon, on which a multi-cluster configuration of the \gls{mlt} accelerator fits together with a 64-bit RISC-V \host \gls{cpu}.
This configuration will not only offer higher accelerator performance but also an open-source soft-macro \host \gls{cpu}, which contrasts with the ``black box'' hard-macro Cortex-A53 host \gls{cpu} of \emph{Aurora} and \emph{Blizzard}.

\begin{table}[htb]
    \centering
    \footnotesize
    \rowcolors{2}{white}{gray!20}
    \renewcommand{\arraystretch}{1.25}
    \begin{tabularx}{\columnwidth}{ X p{.52\columnwidth} c c }
        \toprule
        \textbf{Kernel} & \textbf{Accelerated computation} & \multicolumn{2}{c}{\textbf{Complexity \bigO{}}} \\
        \rowcolor{white} & & \textbf{space} & \textbf{comput.}\hspace*{-.5em} \\
        \midrule
        \texttt{2mm}     & $C\sb{i,j} = \sum\sb{k=1}\sp{N} \alpha A\sb{i,k} B\sb{k,j}$                                                                                           & $N\sp{2}$ & $N\sp{3}$ \\
        \texttt{3mm}     & $E = \texttt{2mm}(A,B) \rightarrow F = \texttt{2mm}(C,D)$ \newline $\rightarrow G = \texttt{2mm}(E,F)$                                                & $N\sp{2}$ & $N\sp{3}$ \\
        \texttt{atax}    & $B\sb{i} = \sum\sb{j=1}\sp{N} A\sb{i,j}X\sb{j}$ \newline $\rightarrow Y\sb{i} = \sum\sb{j=1}\sp{N} A\sb{j,i}B\sb{j}$                                  & $N\sp{2}$ & $N\sp{2}$ \\
        \texttt{bicg}    & $Q\sb{i} = \sum\sb{j=1}\sp{N} A\sb{i,j} P\sb{j}$ \newline $\rightarrow S\sb{j} = \sum\sb{i=1}\sp{N} R\sb{i} A\sb{i,j}$                                & $N\sp{2}$ & $N\sp{2}$ \\
        \texttt{conv2d}  & $B\sb{i,j} = \sum\sb{(k,l)=(-1,-1)}\sp{(1,1)} c\sb{k,l} A\sb{i+k,j+l}$                                                                                & $N\sp{2}$ & $N\sp{2}$ \\
        \texttt{covar}   & $E\sb{j} = \alpha \sum\sb{i=1}\sp{M} D\sb{i,j}$; $D\sb{i,j} \mineq E\sb{j}$; \newline $S\sb{i,j} = S\sb{j,i} = \sum\sb{k=1}\sp{N} D\sb{k,i}D\sb{k,j}$ & $N\sp{2}$ & $N\sp{3}$ \\
        \texttt{darknet} & $C\sb{i,j} =\sum\sb{k=1}\sp{N} \alpha A\sb{i,k} B\sb{k,j}$                                                                                            & $N\sp{2}$ & $N\sp{3}$ \\
        \texttt{gemm}    & $C\sb{i,j} = \beta \left( \sum\sb{k=1}\sp{N} \alpha A\sb{i,k} B\sb{k,j} \right)$                                                                      & $N\sp{2}$ & $N\sp{3}$ \\
        \bottomrule
    \end{tabularx}
    \caption{Evaluated kernels and applications.
        Subscripts denote indices, uppercase letters are variables, and lowercase letters are constants.
        Arrows ($\rightarrow$) denote consecutive offloads.
        Semicolons (;) denote consecutive computations within the same offload.
    }%
    \label{tbl:eval:apps}
\end{table}

The evaluated applications and kernels, listed in \cref{tbl:eval:apps}, represent a wide range of accelerator workloads.
From the Polybench/ACC benchmark suite~\cite{grauer2012auto}, \texttt{2mm}, \texttt{3mm}, \texttt{atax}, \texttt{bicg}, and \texttt{gemm} are linear algebra kernels, \texttt{conv2d} is part of the ``stencil'' domain, and \texttt{covar} is part of the ``datamining'' domain.
Together, these commonly accelerated kernels span a wide range of memory acess patterns and operational intensities.
Additionally, \texttt{darknet} is an end-to-end real-time object detection application that implements the YOLO \gls{cnn}~\cite{redmon2016yolo}.
The data for all applications resides in host-managed shared \gls{dram}.
\texttt{3mm}, \texttt{atax}, \texttt{bicg}, and \texttt{darknet} (one layer at a time) are composed of consecutive offloads, denoted by arrows ($\rightarrow$) in the table; all other kernels consist of a single offload.
All benchmarks are compiled with \texttt{-O3} but no specific optimization flags.
We take the time stamps of each accelerated application on the host, and it thus includes all data transfers and synchronization between host and accelerator.
In all case studies, the accuracy of all results is fully maintained and verified.
In all experiments, the host \gls{cpu} runs Linux~4.19.0 on a root file system generated with Buildroot~2019.02.1, and we compile applications with LLVM~9.0.0 (extended as described in \cref{sec:platform:toolchain}).

\subsection{Application-Level Case Study}%
\label{sec:eval:app}

\noindent
We begin with a case study on the application level.
For each of the applications introduced above, we want to answer the following questions:
How should the local memory of the accelerator be partitioned and data transfers organized so that the run time is dominated by computations on local memory?
What is the speed-up compared to letting the accelerator load and store data directly from off-chip main memory?
How should the application be parallelized over the cores in the accelerator, and what is the speed-up from parallelization?

\begin{figure}[htbp]
  \centering
  \includegraphics[width=\columnwidth]{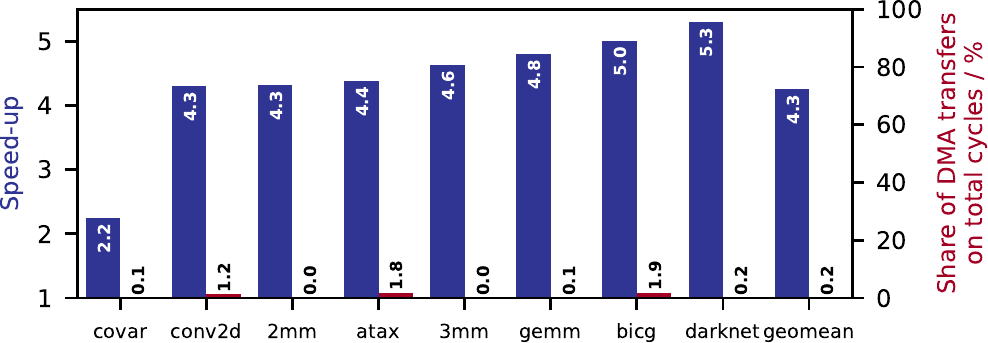}
  \caption{%
    Speed-up of execution on local memory with handwritten \acrshort{dma} transfers compared to execution on external main memory.
    Single accelerator thread.
  }%
  \label{fig:eval:app:dma}
\end{figure}

The first two questions hold the key for making effective use of any accelerator with software-managed local memory.
To answer them, we divide input and output data into tiles.
Assuming all data have the same dimensionality $D$, the side length of one tile is given by $S = \floor{(L / N)\sp{1 / D}}$, where $N$ is the number of data elements (such as different vectors or matrices) and $L$ is the capacity of the L1 for user data in number of words.
With the evaluated accelerator architecture and runtime, $L = \num{28}\,\si{\kibi}$ single-precision (i.e., \SI{4}{\byte}) words can be stored in L1.
Tiling an algorithm is a non-trivial problem to which there is no general solution.
We describe the tiling of one algorithm in the following to give an intuition, and we make the source code of all benchmarks available for full transparency and reproducibility (see link in conclusion).

For the convolutional layers in \darknet, which are implemented as matrix-matrix multiplications, the tile side length of the two input matrices $A$ and $B$ and the output matrix $C$ is $S = 97$.
We loop over the tiles of $A$ and transfer the current tile to L1.
Within that loop, we loop over the tiles of $B$ corresponding to the current horizontal dimension of $A$ and transfer the corresponding tile of $B$ and $C$ in, perform the tiled matrix-matrix multiplication, and transfer the resulting tile of $C$ out.
The other arithmetic kernels are implemented in an analogous manner.
As the left-hand scale in \cref{fig:eval:app:dma} shows, this reduces the run time compared to loading and storing directly from off-chip main memory by \SI{5.3}{\x} for \darknet specifically and by \SI{4.3}{\x} on average\footnote{%
  Whenever we discuss the \emph{average} of normalized numbers, we mean the \emph{geometric mean} (denoted \emph{geomean} in the figures), as reasoned in \cite{fleming1986geomean}.
}.
While this scheme does not exploit double buffering and the nonblocking \gls{dma} transfers that the platform is capable of, the share of cycles spent on \gls{dma} transfers is negligible (max: \SI{1.9}{\percent}, average: \SI{0.2}{\percent}), as the right-hand scale of \cref{fig:eval:app:dma} shows.

Every application lends itself differently to tiling and \gls{dma} transfers:
In applications with high spatial locality, in particular when computation accesses data in the same sequence as it is stored in memory and it does so for large consecutive arrays, the \gls{dma} engine can transfer long continuous data bursts.
This is particularly common in linear algebra and \gls{cnn} kernels: the kernels with the highest speed-up in \cref{fig:eval:app:dma} are all from those domains.

In applications with low spatial locality or divergent access patterns, \gls{dma} transfers are substantially shorter and thus offer lower speed-up.
Nonetheless, the \gls{dma} engine's capability for gather-scatter transfers and many outstanding requests offers a speed-up of more than \SI{4}{\x} even with low spatial locality.
Temporal locality, on the other hand, has an even bigger impact:
For some applications, tiling necessitates that each data element is loaded multiple times because local memory is not large enough to hold all data elements between two use instants.
\texttt{covar} is an example of such an application, where each element of the data matrix has to be loaded twice (once during mean calculation and once while computing the covariance matrix).
This reload factor of two reduces the speed-up by \gls{dma} transfers by almost \SI{2}{\x} to only \SI{2.2}{\x}.

\begin{figure}[htbp]
  \centering
  \includegraphics[width=\columnwidth]{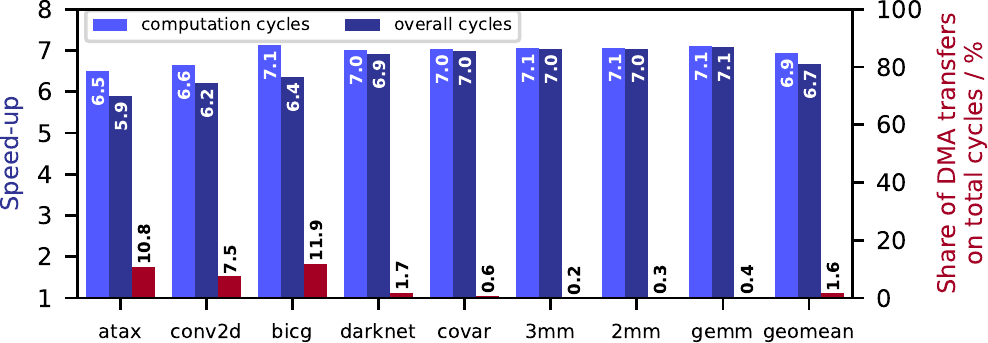}
  \caption{%
    Speed-up of execution with 8 accelerator threads compared to execution with 1 accelerator thread.
    Execution on local memory with handwritten \acrshort{dma} transfers.
  }%
  \label{fig:eval:app:parallel}
\end{figure}

The third question -- how an application should be parallelized -- holds the key for making effective use of any parallel accelerator.
\hero{}'s OpenMP runtime library enables to answer this question efficiently be experimentation:
For the computation on one tile, we simply annotate the outermost computational loop with \texttt{\#pragma omp for} to distribute its execution over the cores of the accelerator.
As the left bar for each application in \cref{fig:eval:app:parallel} shows, this reduces the computation cycles by \SIrange{6.5}{7.1}{\x} (average: \SI{6.9}{\x}) on an 8-core cluster.
Even higher speed-ups by parallelization could be achieved by optimizing the loop schedule and stride, but we set that aside because it is sensitive to data size.
The overall application speed-up by parallelization, shown by the middle bar for each application, is between \SIrange{5.9}{7.1}{\x} (average: \SI{6.7}{\x}).
The right bar shows why the computation-only speed-up cannot be achieved for the overall application:
The \gls{dma} transfers are not sped up by parallelization, so their share on the total cycles increases by the overall speed-up factor.
Due to Amdahl's law, this limits the overall speed-up achievable by parallelization.
On average, \SI{2.2}{\percent} cycles spent on \gls{dma} transfers result in a modest decrease from \SI{7.0}{\x} to \SI{6.6}{\x}.
However, for some applications, such as \texttt{covar}, \SI{10.3}{\percent} cycles spent on \gls{dma} transfers reduce the parallelization speed-up from \SI{7.4}{\x} to \SI{6.6}{\x}.
This may justify a more complex double-buffered implementation of an application.

This benchmark analysis shows how \hero's full-stack hardware and software allows to rapidly explore and optimize the accelerated performance of domain-relevant applications on a heterogeneous computer prototype:
The high emulation throughput allows to study realistic problem sizes, and the complete software stack allows to adapt and tune real-world applications and representative kernels with reasonable effort and make informed optimization decisions.
Furthermore, the fully open hardware implementation allows tracing and profiling hardware, as well as optimizing it.

\subsection{Runtime-Level and Toolchain Case Study}%
\label{sec:eval:toolchain}

\noindent
Tiling an algorithm for efficient execution on accelerator-local memory is not only an intellectual effort but also requires extra code to be written, verified, and maintained.
\hero{}'s \gls{api} is designed to simplify this task for device-specific operations such as \gls{dma} transfers, which can be executed with a single function call, and fork-join parallelism, which is available through the standardized OpenMP pragmas that call into the runtime library.
However, the part of tiling that is specific to each algorithm cannot be substantially simplified by a runtime library.

\begin{figure}[htbp]
  \centering
  \includegraphics[width=\columnwidth]{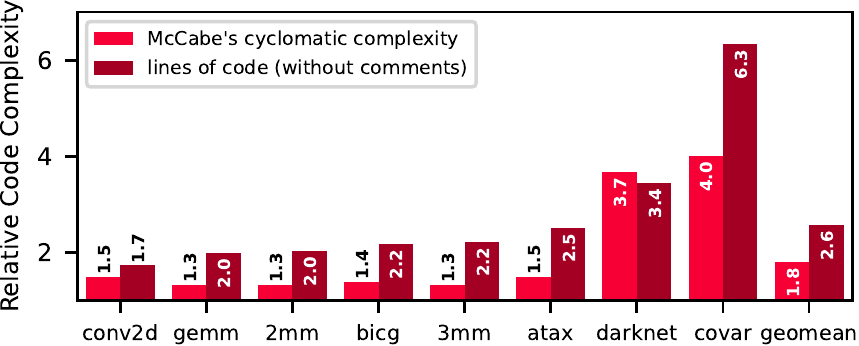}
  \caption{%
    Code complexity with handwritten tiling and \acrshort{dma} transfers compared to the unmodified code of each application.
    The light red bars show McCabe's cyclomatic complexity and the dark red bars show lines of code without comments.
  }%
  \label{fig:eval:app:complexity}
\end{figure}

The code complexity increase by handwritten tiling and \gls{dma} transfers compared to unmodified code is shown in \cref{fig:eval:app:complexity} for each application.
We used the CCCC tool\footnote{\url{https://sourceforge.net/projects/cccc}}~\cite{littlefair2001code} on the accelerated part of each application and extracted two of its results:
(1) The lines of code (without comments), which can be an indication for the effort of writing and reading a piece of code.
(2) McCabe's cyclomatic complexity, which counts the number of linearly independent paths through a piece of code, and which can be an indication for the effort of understanding and verifying a piece of code.
The results show three coarse categories of applications:
First, the six applications on the left are tiled in a single dimension, which is a modest effort: the lines of code increase by \SIrange{1.7}{2.5}{\x} and the cyclomatic complexity increases by \SIrange{1.3}{1.5}{\x}.
On average, the lines of code overhead by 1D tiling is \SI{2}{\x} and the cyclomatic complexity increase is \SI{1.4}{\x}.
Second, \texttt{darknet} with its \gls{cnn} layers is implemented with two-dimensional tiling and \gls{dma} transfers.
2D tiling substantially increases both the cyclomatic complexity (\SI{3.7}{\x}) and the lines of code (\SI{3.4}{\x}).
Third, \texttt{covar} is also implemented with 2D tiling, but the implementation is additionally split over two separate iterations through the entire data.
This means the ca.\ \SI{3}{\x} lines of code overhead by 2D tiling incurs twice, leading to a total \SI{6.3}{\x} lines of code overhead, while the cyclomatic complexity increases by the same factor as for \texttt{darknet}.
In summary, the additional effort and maintenance cost for tiling an algorithm ranges from modest (\SI{1.7}{\x} LOC, \SI{1.5}{\x} cyclo.\ compl.) to very high (\SI{6.3}{\x} LOC, \SI{4.0}{\x} cyclo.\ compl.) and is certainly not negligible on average (\SI{2.6}{\x} LOC, \SI{1.8}{\x} cyclo.\ compl.).

OpenMP assumes a cache-based memory hierarchy, leading to low performance on \gls{spm}-based memory hierarchies if a program is not manually tiled.
To save these substantial manual tiling efforts, an optimal solution would be if the toolchain could automatically transform the untiled algorithm code to manage the memory hierarchy.
The \emph{AutoDMA} feature, introduced in \cref{sec:platform:toolchain}, brings this to \hero{}.
Effectively, this means that the software-managed memory hierarchy of \hero{} can be programmed as easily as a cache-based system.

\begin{figure}[htbp]
  \centering
  \includegraphics[width=\columnwidth]{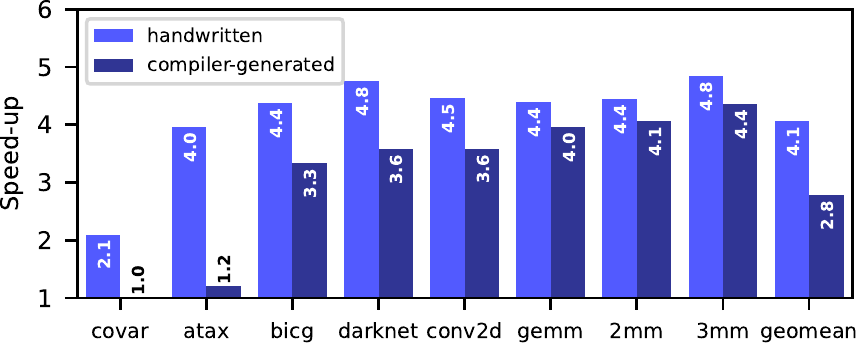}
  \caption{%
    Speed-up of execution on local memory with \emph{compiler-generated} tiling and \acrshort{dma} transfers compared to execution on external main memory.
    8 accelerator threads.
    The light blue bars show the speed-up of the handwritten implementation for comparison; the dark blue bars show the speed-up achieved by the compiler.
  }%
  \label{fig:eval:app:hc}
\end{figure}

The speed-up of compiler-generated and handwritten tiled code over unmodified OpenMP code is shown in \cref{fig:eval:app:hc}.
While the handwritten tiled code has a significantly higher complexity than the unmodified OpenMP code, as shown in \cref{fig:eval:app:complexity}, compiler-generated tiling requires zero code changes.
The benchmarks in \cref{fig:eval:app:hc} can be divided into two categories:
For \texttt{covar} and \texttt{atax}, the speed-up achieved by the compiler is marginal.
For all other benchmarks, the speed-up achieved by the compiler is comparable to that of handwritten code.
The benchmarks in the latter category feature large segments of contiguous memory accesses (spatial locality), and achieve on average \SI{85}{\percent} of the speed-up of handwritten code.
The remaining \SI{15}{\percent} come from leveraging programmer insights (i.e., information not expressed in the code) to reduce the number of reconfigurations of the \gls{dma} engine:
The handwritten code transfers multiple rows of matrices at once, possible by the understanding that the first element of row $N+1$ is next in memory to the last element of row $N$.
The compiler was not able to reconstruct this information, due to \emph{array-to-pointer decay} in which the dimensions of data structures are lost.
Without this information, the compiler considers multiple rows as non-contiguous and initiates a new \gls{dma} burst for each row, which adds an overhead compared to the single \gls{dma} burst in the handwritten code.
Nonetheless, AutoDMA provides a speed-up of up to \SI{4.4}{\x} without any code changes.
Expert programmers still have the option to turn this feature off and implement tiling manually to extract the last tens of percent of performance.

For two benchmarks (\texttt{covar} and \texttt{atax}), the compiler-generated code cannot compete with the handwritten code.
This can also be attributed to memory access patterns: a significant part of memory accesses are performed column-wise, i.e., in non-contiguous blocks.
This effect is aggravated by the tile shape selected by the compiler, which inadvertently maximizes the number of column-wise accesses per tile, rather than contiguous row-wise accesses.
This is due to the loop ordering of the benchmarks, which the AutoDMA feature does not rewrite\footnote{%
  Tools that reorder loops, such as polyhedral analyses and transformations~\cite{grosser2012polly}, could be used to preprocess the code, or the benchmarks could be manually rewritten using classical spatial locality optimizations.
}.
Spatial locality is also important for performance on cache-based systems, but the issue is aggravated on \hero{} where the \gls{dma} engine in this case is used to transfer individual words.
As such, it is an extreme case of the overhead discussed for the previous category, where the compiler could not find sufficiently large chunks of contiguous memory.
Despite these problems, the performance with AutoDMA is on-par to up to \SI{20}{\percent} higher than the OpenMP baseline, due to the high bandwidth of the \gls{dma} engine.

In summary, for programs with high spatial locality, \hero{}'s AutoDMA feature provides performance comparable to handwritten code, without the need for explicit tiling and \gls{dma} transfers.
This reduces the execution time of unmodified OpenMP programs by up to \SI{4.4}{\x} on software-managed memory hierarchies, achieving \SI{85}{\percent} of the speed-up of handwritten code.
This makes software-manged memory hierarchies as easy to program as their hardware-cache-based counterparts.
Similarly to hardware-managed caches, AutoDMA provides no significant improvements for programs with low spatial locality.

\hero{} is a unique platform to analyze, develop, and optimize such compiler and runtime techniques, because it allows executing real applications and reference benchmarks on the actual \gls{rtl} logic of a heterogeneous \gls{soc}, and because all its hardware and software components are open-source and permissively licensed.

\subsection{System Architecture-Level Case Study}%
\label{sec:eval:system}

\noindent
Our third case study examines the impact of an architectural design decision: How does the data width of the accelerator into the shared interconnect and main memory influence the performance of accelerated applications?
To answer this question, we customize the on-chip network of the accelerator once to half the data width (\SI{32}{\bit}) and once to twice the data width (\SI{128}{\bit}) and remeasure our applications.

\begin{figure*}[tbp]
  \centering
  \ifthesis%
    \includegraphics[width=15cm,angle=90]{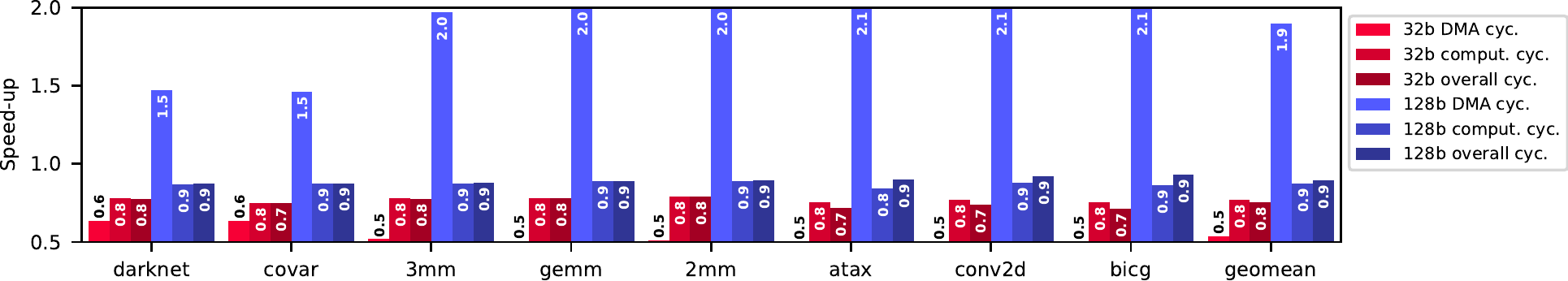}%
  \else%
    \includegraphics[width=\textwidth]{eval_speedup_datawidth.pdf}%
  \fi%
  \caption{%
    Speed-up of execution with an accelerator on-chip network data width of \SI{32}{\bit} or \SI{128}{\bit} compared to the default of \SI{64}{\bit}.
    Each application has six bars:
    The three bars on the left are for \SI{64}{\bit} data width, the three bars on the right are for \SI{128}{\bit} data width.
    Of each three bars, the left shows the speed-up of only the \acrshort{dma} cycles, the middle of only the computation cycles, and the right of the total cycles.
  }%
  \label{fig:eval:datawidth}
\end{figure*}

\Cref{fig:eval:datawidth} shows the speed-up (for values $> 1$) or slow-down (for values $< 1$) for an accelerator on-chip network data width of \SI{32}{\bit} (left three bars of each application) and \SI{128}{\bit} (right three bars) compared to \SI{64}{\bit}.
The leftmost bar in each group of three bars compares the cycles spent on \gls{dma} transfers:
For most applications, halving the data width of the on-chip network results in a speed-up of \SI{0.5}{\x}, and doubling the data width results in a speed-up of \SI{2}{\x}, as expected.
The exception, however, is \texttt{darknet}, with \SI{0.6}{\x} for half the data width and \SI{1.5}{\x} for double the data width.
\texttt{covar} and \texttt{darknet} are the only application to use two-dimensional \gls{dma} transfers, which are composed of many relatively short bursts.
This transfer pattern does not fully saturate the given on-chip network, which results in lower speed-ups for wider data widths.
That is an important insight for optimizing the on-chip network if \gls{dma} performance was critial for application performance.
However, as we know from the application-level case study (\cref{sec:eval:app}), \gls{dma} transfers only account for at most \SI{11.9}{\percent} (average: \SI{2.4}{\percent}) of the application cycles.
The majority of cycles is spent in computations, and the middle of each three bars compares the cycles spent on computations:
Surprisingly, the data width of the on-chip network also has a significant impact on them.
For \SI{32}{\bit}, the fetch bandwidth of instructions into the cache is halved, which leads to more instruction fetch stall cycles and reduces computational performance.
For \SI{128}{\bit}, the fetch bandwidth for instructions could be doubled, but the instruction cache can only fetch at most \SI{64}{\bit} per cycle, so that has no impact.
To accommodate the wider memory interface of the \gls{dma} engine, the \gls{tcdm} interconnect in the accelerator cluster has to be changed from $14 \times 16$ to $18 \times 32$.
This configuration causes on average \SI{15}{\percent} more contention on the \gls{tcdm} despite the higher number of banks.
A careful realignment of the cores on the \gls{tcdm} interconnect could alleviate this, but the gist is that a wider accelerator on-chip network does not automatically increase performance.
In fact, as the rightmost bar of each application shows, application performance decreases by \SI{10}{\percent} on average if the design of the cluster is not simultaneously adapted.

Such insights from fully measured application executions are central for making substantiated decisions on the system architecture and for prioritizing engineering efforts.
The closer the measured prototype is to the final design, the higher the quality of the measurements.
Effects such as those discussed in this section would be extremely difficult to model with a simulator, as they depend on fine-grained interaction between several hardware components.
Capturing this interaction quantitatively with non-cycle-accurate architectural simulation is a very intricate and error-prone task.
Thus, an application-programmable heterogeneous research platform with a complete hardware and software stack, such as \hero{}, is a key enabler for architecture-level performance exploration.

\subsection{Accelerator ISA-Level Case Study}%
\label{sec:eval:accel}

\noindent
Specialized instructions are an important part of many domain-specific accelerators.
They are often designed and evaluated in an \gls{iss} or in \gls{rtl} simulations.
The drawback of \gls{iss} is that it is inaccurate as performance model because it does not capture microarchitectural effects.
\Gls{rtl} simulation models the microarchitecture accurately, but it is only feasible for small data set and does not take communication outside the accelerator, which influences the memory subsystem and thereby the execution of the accelerated kernel, into account.
Thus, a heterogeneous research platform is required to quantify the impact of specialized accelerator \gls{isa} extensions in heterogeneous computing with real-world data sets.

In this case study, we answer the question ``How much do instructions from the \texttt{Xpulpv2} \gls{isa} extension speed up execution of heterogeneous applications compared to the standard \texttt{rv32imafc} \gls{isa}?''.
As described in \cref{sec:platform:toolchain}, we have extended the RISC-V LLVM backend to automatically emit \texttt{Xpulpv2} instructions during machine code generation.
The evaluated kernels process data at full precision (i.e., 32-bit integers or floats) and therefore cannot make use of the quarter- or half-precision packed \gls{simd} instructions, which would offer a significant speed-up for reduced-precision processing.

\begin{figure}[htbp]
  \centering
  \includegraphics[width=\columnwidth]{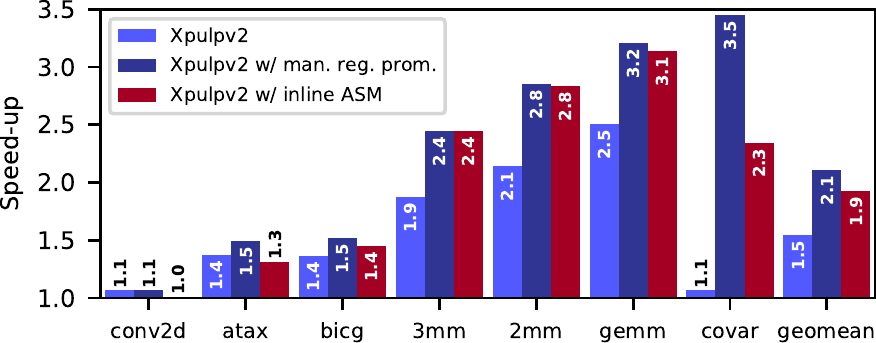}
  \caption{%
    Speed-up of execution with custom instructions (\texttt{Xpulpv2}) compared to the standard RISC-V (\texttt{RV32IMAFC}) \acrshort{isa}. Execution on local memory with handwritten \acrshort{dma} transfers and 8 accelerator threads.
    For each application,
    the first bar shows the speed-up by \texttt{Xpulpv2} instructions without manual register promotion,
    the second bar by \texttt{Xpulpv2} instructions with manual register promotion,
    and the third bar by implementing the innermost loop with inline assembly instructions (including manual register promotion and \texttt{Xpulpv2} instructions).
  }%
  \label{fig:eval:app:accel}
\end{figure}

The speed-up of the \texttt{Xpulpv2} \gls{isa} extension over the standard RISC-V \texttt{RV32IMAFC} \gls{isa} is shown in \cref{fig:eval:app:accel}.
We measure the total accelerator cycles with handwritten \gls{dma} transfers and 8 accelerator threads.
As the first bar of each application shows, simply enabling \texttt{Xpulpv2} provides a speed-up of \SI{1.5}{\x} on average.
Starting with \texttt{gemm} as an example, we find that the compiler replaces the inner two compute loops by hardware loops.
This is optimal, as there is only hardware for two loops.
The body of the innermost loop is halved from 10 instructions (2 loads, 4 additions, 2 multiplications, 1 store, and 1 branch) to 5 instructions (2 post-increment loads, 1 multiplication, 1 \gls{mac}, and 1 store), while the bodies of the outer levels stay mostly identical.
Apart from the store, which could be hoisted out of the innermost loop by a memory-to-register optimization pass, the innermost loop is optimal\footnote{%
  For \texttt{gemm}, the multiplication by $\alpha$ could be hoisted out of the innermost loop for all data types where multiplication is distributive over addition.
  However, this is an algebraic transformation and does not apply to all data types (such as \texttt{float}s), so we do not consider it.
}, and it is also optimally scheduled.
The resulting speed-up of \SI{2.5}{\x} can be attributed to halving the instructions within the innermost loop (ca.\ \SI{2}{\x} speed-up) and hardware loops as well as less instructions in the outer loops (ca.\ \SI{0.5}{\x} speed-up).
Manually hoisting the store out of the innermost loop significantly improves performance further.
Again looking at \texttt{gemm}, this reduces the innermost loop from 5 to 4 instructions, and the resulting relative speed-up of \SI{1.28}{\x} is aligned with the reduction in instructions.
The same findings hold for \texttt{3mm} and \texttt{2mm}, and a comparison with an inline assembly implementation of the innermost loop reveals that the instructions generated by the compiler perform on-par or better than the expert-written instructions.
However, some benchmarks behave quite differently:
For \texttt{conv2d}, \texttt{atax}, and \texttt{bicg}, the \texttt{Xpulpv2} \gls{isa} extension provides only between \SIrange{10}{50}{\percent} of speed-up -- both with compiler-generated instructions and with an expert-written inner loop body.
There are two main reasons for this:
First, the kernels are not as well suited for post-increment memory accesses as the matrix-matrix multiplication kernels.
For \texttt{atax}, the increment of one of the two loads in the innermost loop is too large to be used in post-increment.
For \texttt{conv2d}, the 2D ($3 \times 3$) loads in the innermost loop leave some opportunities for post-increment loads, but they are complex to exploit.
Even the expert-written instructions, which use as many post-increment accesses as possible, do not bring a significant speed-up.
Second, hardware loops are not inferred for the innermost loops.
This could be because the innermost loop iterates over the rows in a tile, and the number of rows changes depending on the tile index.
This does not fundamentally preclude the use of hardware loops, however, so it is a current compiler limitation.
Finally, \texttt{covar} sees a very high speed-up with \texttt{Xpulpv2}, but only with manual memory-to-register promotion.
This simple change in the code enables the compiler to infer a hardware loop.
The instructions generated by the compiler substantially outperform the expert-written inner loop, due to better scheduling.

In summary, the \texttt{Xpulpv2} \gls{isa} extension has the potential to significantly accelerate all kernels we evaluated, mainly through post-increment memory accesses and hardware loops.
Especially the latter are not trivial for the compiler to generate in all cases, however, which currently leads to speed-ups between \SIrange{1.1}{3.5}{\x} (average: \SI{2.1}{\x}).
While the impact of changes to the accelerator \gls{isa} could also be studied in isolation (e.g., in \gls{rtl} simulation), evaluating within a heterogeneous prototype system (such as \hero{}) produces more representative results, because the balance and interaction between memory transfers from and to the shared main memory and computation are taken into account, and because a prototype running at tens of \si{\mega\hertz} makes it feasible to work with real-scale data sets.

\section{Related Work}%
\label{sec:related_work}

\def \chipyardlogo {\raisebox{-2pt}{\includegraphics[height=10pt]{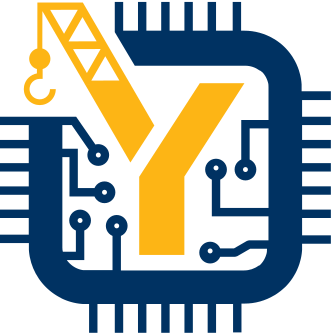}}}
\def \ohglogo {\raisebox{-2pt}{\includegraphics[height=10pt]{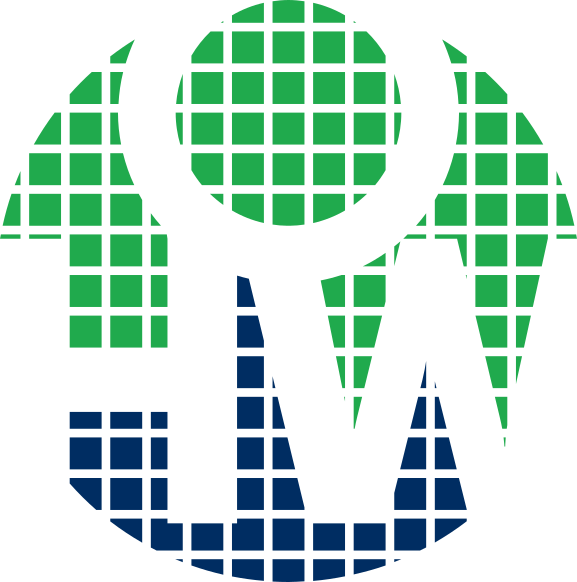}}}
\def \pulplogo {\raisebox{-2pt}{\includegraphics[height=10pt]{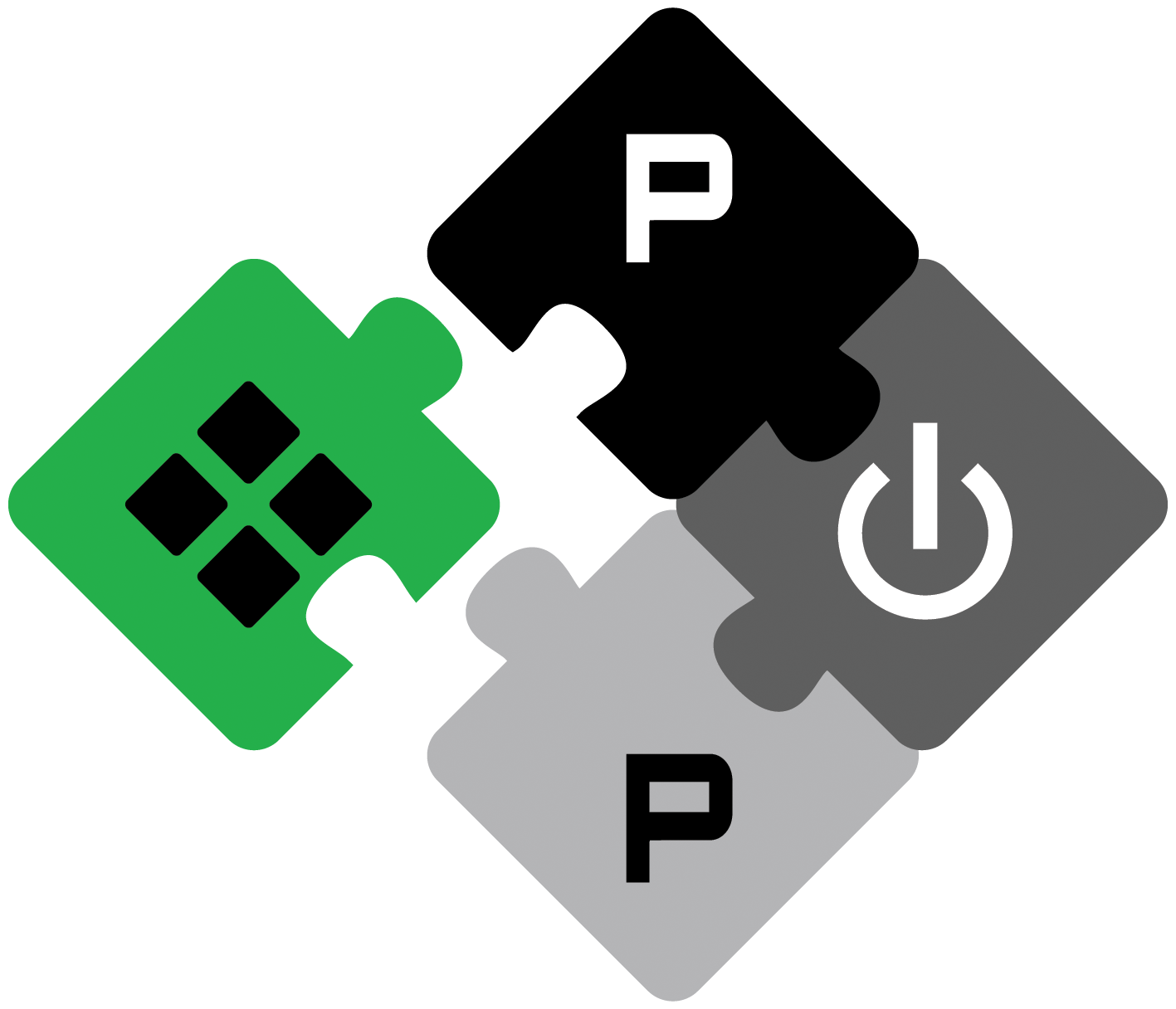}}}
\def \snitchlogo {\raisebox{-2pt}{\includegraphics[height=10pt]{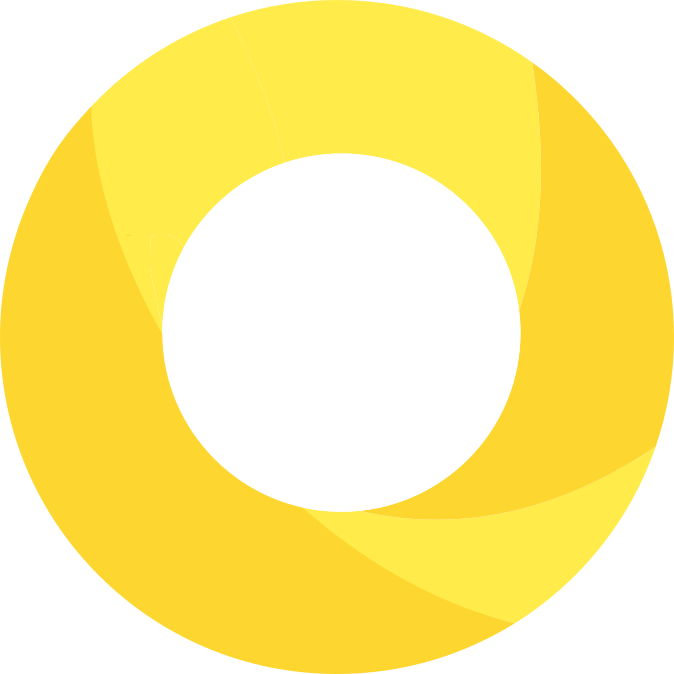}}}
\def \sameline #1#2{\rule[2pt]{#1}{.25pt}~#2~\rule[2pt]{#1}{.25pt}}
\def \progmodel #1{\raisebox{0pt}{\includegraphics[height=8pt]{prog_model_scale_#1.png}}}

\newcommand\circledsym[2]{%
  \adjustbox{height=1.15em,margin*=0 -1.25 -2.5 0}{%
    \tikz\node[circle,color=white,fill=#1,inner sep=.2pt,font=\bfseries]{#2};%
  }
}
\newcommand{\goodyes}{\circledsym{OliveGreen}{$\pmb\checkmark$}}
\newcommand{\badno}{\circledsym{OrangeRed}{\textsf{X}}}

\def \citefiresim {\cite{karandikar2018firesim,karandikar2020fireperf}}
\def \citehero {\cite{kurth2017herov1,kurth2018herov1}}
\def \citeopenesp {\cite{mantovani2020openesp,giri2021openesp}}

\begin{table*}[htbp]
  \centering
  \maxsizebox{\textwidth}{!}{%
  \begin{tabular}{ l c c c c c c c c c }
    \toprule
    \textbf{Platform} & \textbf{FAME} & \textbf{Emul.} & \textbf{HW} & \textbf{Open-} & \multicolumn{5}{c}{\textbf{Heterogeneous SoC Features}} \\
    & & \textbf{Focus} & \textbf{Ecosys.} & \textbf{Source} & \scriptsize\textbf{Hosts} & \scriptsize\textbf{Accelerators} & \scriptsize\textbf{Program.\ Model} & \scriptsize\textbf{Accel.\ Coh.} & \scriptsize\textbf{MDM} \\
    \midrule
    Cadence Palladium~{\scriptsize\cite{cadence2021palladium}} & 0 & Dig & \textcopyright{} & \badno{} & \multicolumn{5}{c}{\sameline{30ex}{customer-provided}} \\
    Siemens Veloce Strato~{\scriptsize\cite{siemens2021veloce}} & 0 & Dig & \textcopyright{} & \badno{} & \multicolumn{5}{c}{\sameline{30ex}{customer-provided}} \\
    Synopsys ZeBu~{\scriptsize\cite{synopsys2021zebu}} & 0 & Dig & \textcopyright{} & \badno{} & \multicolumn{5}{c}{\sameline{30ex}{customer-provided}} \\
    OpenPiton~{\scriptsize\cite{balkind2016openpiton,balkind2020openpiton,balkind2020byoc}} & 0 & Many & p & \goodyes{} & \scriptsize{T1, CVA6, PRV32, ao486} & MIAOW, NVDLA & \progmodel{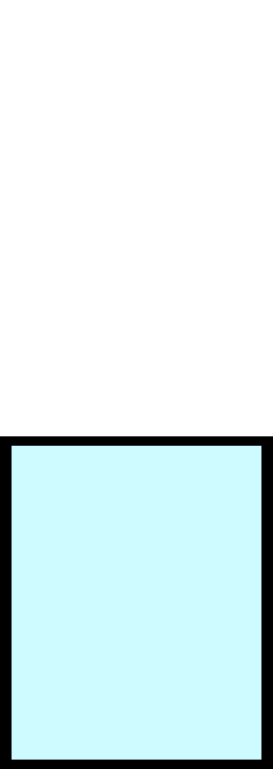} & 4 & \badno{} \\
    MEG~{\scriptsize\cite{zhang2020meg}} & 0 & Mem & p & \goodyes{} & Boom & custom & n/d & n/d & \badno{} \\
    DART~{\scriptsize\cite{wang2014dart,prasad2021fpganoc}}, DuCNoC~{\scriptsize\cite{kamali2018ducnoc}} & 3 & Net & d & \badno{} & \multicolumn{3}{c}{\sameline{26.5ex}{n/a}} & n/d & \badno{} \\
    FireSim~{\scriptsize\citefiresim{}} & 1, 5 & Dig & \chipyardlogo & \goodyes{} & Rocket & NVDLA, HLS & \progmodel{1} & n/d & \badno{} \\
    Centrifuge~{\scriptsize\cite{huang2019centrifuge}} & 1, 5 & HeSoC & \chipyardlogo & \goodyes{} & Rocket & HLS & \progmodel{1} & 2, 4 & \badno{} \\
    OpenESP~{\scriptsize\citeopenesp{}} & 0 & HeSoC & p & \goodyes{} & CVA6, LEON3 & custom, HLS & \progmodel{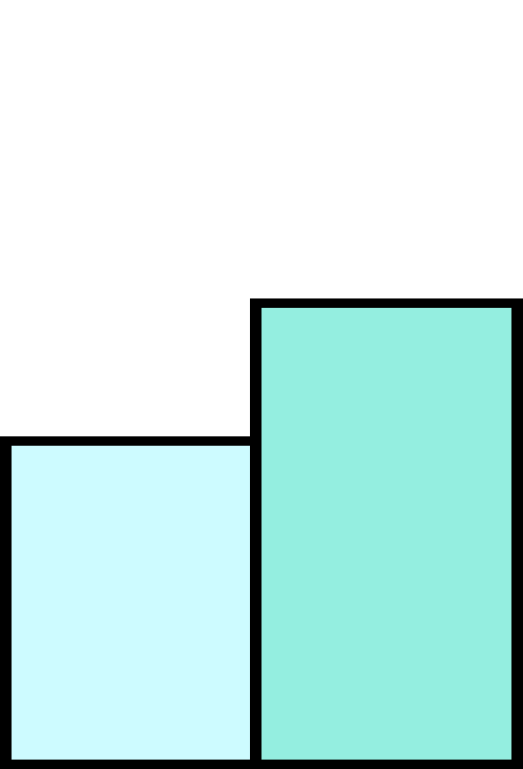} & 1, 2, 4 & \badno{} \\
    HERO~{\scriptsize\citehero{}} & 0 & HeSoC & \pulplogo{}, \ohglogo{} & \goodyes{} & A9 & \pulplogo{} & \progmodel{1}, \progmodel{2}, \progmodel{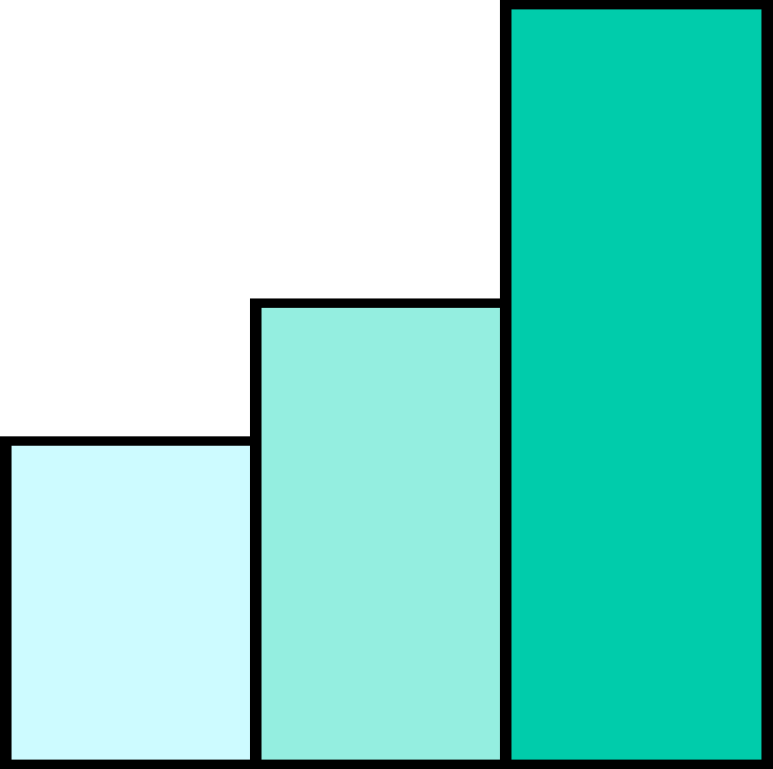} & 1, 2 & \badno{} \\
    HEROv2 {\scriptsize[this work]} & 0 & HeSoC & \pulplogo{}, \ohglogo{} & \goodyes{} & A53, CVA6 & \pulplogo{}, \snitchlogo{} & \progmodel{1}, \progmodel{2}, \progmodel{3} & 1, 2, 3 & \goodyes{} \\
    \bottomrule
  \end{tabular}
  }
  \caption{%
    Comparison of computer emulation systems on programmable logic devices.
    Legend:
    \textbf{FAME}: taxonomy numbers~\cite{zhangxi2010fame} |
    \textbf{Emul.\ Focus}: \uline{Dig}ital hardware, \uline{He}terogeneous \uline{SoC}s, \uline{Many}cores, Near-\uline{Mem}ory Processing, On-Chip \uline{Net}works |
    \textbf{HW Ecosys.}: \textcopyright{} commercial, as \uline{p}orted or \uline{d}eveloped by platform maintainers, \pulplogo{} PULP, \ohglogo{} OpenHW Group, \chipyardlogo{} Chipyard |
    \textbf{Hosts}: ARM Cortex-\uline{A9} or -\uline{A53}, OpenHW Group \uline{CVA6}, OpenSPARC \uline{T1}, Berkeley out-of-order machine (\uline{Boom}), Gaisler \uline{LEON3}, \uline{P}ico\uline{RV32} |
    \textbf{Accelerators}: \pulplogo{} PULP cluster, \snitchlogo{} Snitch cluster, \uline{MIAOW} GPGPU, \uline{custom} logic, \uline{NVDLA}, \uline{HLS}-generated |
    \textbf{Programming Model}: accelerator separately programmable through \progmodel{1} an OS driver or \progmodel{2} a user-space API, or \progmodel{3} unified host+accelerator programming with heterogeneous OpenMP applications, or \uline{n/d} not defined | %
    \textbf{Accelerator Coherence Modes}~\cite{zuckerman2021cohmeleon}: \uline{1} non-coherent DMA, \uline{2} LLC-coherent DMA, \uline{3} coherent DMA, \uline{4} coherent cache, or not defined (\uline{n/d}) |
    \textbf{MDM}: mixed-data-model (e.g., 64-bit host + 32-bit accelerator) programming supported.
  }%
  \label{tbl:related_work}
\end{table*}

\noindent
Emulation systems on \glspl{fpga} or custom programmable logic are widely used to get cycle-accurate results at a clock frequency of multiple \si{\mega\hertz} and turn-around times of few hours to days.
In the \gls{fame} taxonomy~\cite{zhangxi2010fame}, \hero{} is a \emph{Direct \gls{fame}} system, which are characterized by implementing the target system with a one-to-one correspondence in clock cycles on an \gls{fpga}.
Commercial Direct \gls{fame} systems include Cadence Palladium~\cite{cadence2021palladium}, Siemens Veloce~Strato~\cite{siemens2021veloce}, and Synopsys ZeBu~\cite{synopsys2021zebu}.
Those systems are capable of emulating up to 20 billion \gls{asic} \glspl{ge} at up to \SI{10}{\mega\hertz} and can cost millions of USD.
\hero{} can scale over multiple \glspl{fpga} with chip-to-chip \gls{fmc} connections, which are supported by all of \hero{}'s carrier silicon.
Depending on the design, the system interconnect, the accelerator interconnect, or both can extend over multiple \glspl{fpga} through \gls{fmc} and QSFP+ connections.
\hero{}'s currently largest carrier silicon, Xilinx' VCU128, offers ca.\ 40 million \gls{asic} \glspl{ge} and can communicate with other \glspl{fpga} at more than \SI{650}{\giga\bit\per\second}.
Depending on \gls{fpga} and configuration, \hero{}'s clock frequency is between \SIrange{20}{100}{\mega\hertz}.

\Gls{fpga}-based computer system emulators are common in industry and research.
The following recent works are comparable with ours (see \cref{tbl:related_work} for an overview and \cite{angepat2014fpgaemul} for a broader survey of older approaches up to 2014):
OpenPiton~\cite{balkind2016openpiton} is an open-source many-core research framework that can be implemented on an \gls{fpga}.
It comes with a cache-coherent on-chip network and by now supports four different processor cores~\cite{balkind2020byoc,balkind2020openpiton}, among them CVA6 also supported by \hero{}.
The most recent version of OpenPiton optionally includes an open-source GPGPU or Nvidia's deep learning accelerator (NVDLA), which can be programmed from Linux running on the processor cores.
This recent developments allow using OpenPiton for research on heterogeneous computing, which is \hero{}'s focus, but the full hardware-software stack integration of accelerators, from \gls{api} to accelerator-specific compiler backend, remains \hero{}'s distinguishing feature.
MEG~\cite{zhang2020meg} is a system emulation infrastructure for near-data processing implemented on an \gls{fpga}.
It features four 64-bit RISC-V Boom cores as host processor and a near-memory accelerator whose architecture and \gls{isa} are not specified.
Like \hero{}, MEG features a Linux-booting host processor and is implemented on a VU37P, but unlike \hero{}, the focus is on near-memory accelerators that seem to have a fixed function, as accelerator programming, memory hierarchy, data transfers, and communication with the host are not discussed.
DART~\cite{wang2014dart} accelerates the simulation of on-chip networks by mapping them onto an \gls{fpga}.
It provides programmability by decoupling the simulator architecture from the architecture of the simulated on-chip network.
Similarly, DuCNoC~\cite{kamali2018ducnoc} maps on-chip networks to the \gls{pl} of a Zynq-7000 \gls{soc}.
Like \hero{}, the on-chip network is highly configurable and modeled cycle-accurately at \SI{10}{\mega\hertz} and more in DART and DuCNoC, but unlike \hero{}, the remainder of the computer system remains in a higher-level simulator that injects traffic into the on-chip network.
Prasad~\textit{et.\,al.}~\cite{prasad2021fpganoc} improve on DART by specializing the microarchitecture of on-chip network components to the target \gls{fpga} architecture, which reduces the required hardware resources by \SI{70}{\percent} and the average packet latency by \SI{20}{\percent}.
In contrast, \hero{}'s components are not specialized to \glspl{fpga}, which means they consume more hardware resources than minimally required but also that they match an \gls{asic} implementation cycle-by-cycle.

FireSim~\citefiresim{} extends \gls{fpga}-based emulation to Amazon~EC2~F1, a public cloud \gls{fpga} platform.
On the \gls{fpga} of each instance, FireSim allows instantiating modules from the Chipyard~\cite{amid2020chipyard} ecosystem (e.g., the 64-bit RISC-V Rocket core, a L2 cache, a \gls{nic}, and fixed-function accelerators such as the Hwacha vector processor).
Multiple instances are connected over the datacenter network and C++ simulation models to emulate datacenter clusters with multiple server nodes.
Like \hero{}, FireSim comes with an OS-capable multi-core CPU, but unlike \hero{}, the focus is on datacenter clusters and networking instead of heterogeneous computing with different \glspl{isa}, data models, execution models, and memory subsystems.
Centrifuge~\cite{huang2019centrifuge} extends FireSim with a flow that generates heterogeneous \glspl{soc} containing user-defined HLS accelerators together with a Linux driver for them.
In contrast, accelerators in \hero{} can be interfaced with user-space libraries or in heterogeneous OpenMP applications, but the accelerator software is not auto-generated.

Research platforms that combine HW and SW components are less common.
OpenESP~\citeopenesp{} is a research platform for heterogeneous \gls{soc} design.
It provides a methodology and components to integrate processors (among them CVA6 also supported by \hero{}) and HLS-generated accelerators with a 2D-mesh on-chip network.
Like in \hero{}, the accelerators have a \gls{dma} engine and can share virtual addresses with a processor through an \gls{iommu} and a Linux driver.
Unlike in \hero{}, accelerators are not programmable with a full-featured standard \gls{isa}, and there is thus no OpenMP offloading support and no heterogeneous \gls{api}, runtime libraries, and toolchain that span across host processors and accelerators.
HEROv1~\citehero{} does provide the components that enable the evaluation of heterogeneous applications on a mixed-\gls{isa} computer, but its toolchain is fundamentally limited to 32-bit hosts and accelerators~\cite{kurth2020cc}.
Additionally, it has no \gls{api} that unifies programming over multiple accelerators; it features one host and one accelerator architecture, and hardware and software are tailored to those instead of being modular; and its on-chip network is limited to simple configurations (e.g., fixed 64-bit data width) and topologies (e.g., central crossbar), which do not meet the demands of modern heterogeneous computers.

Accelerators have been designed specifically for \glspl{fpga}.
GRVI~Phalanx~\cite{gray2016grviphalanx} 32-bit RISC-V soft processor array that scales to more than 1000 cores on a Xilinx VU9P \gls{fpga}.
2GRVI~Phalanx~\cite{gray20192grviphalanx} extends that to more than 1000 64-bit RISC-V cores on a Xilinx VU37P.
The DRAGON architecture~\cite{abdelhamid2021dragon} is a 64-bit custom-\gls{isa} cluster-based multiprocessor that scales to 144 cores on a Xilinx VU37P.
In contrast, the accelerator in \hero{} is not specialized for \glspl{fpga} but has identical \gls{rtl} code as for \gls{asic} tapeouts.
Its components, from cores~\cite{gautschi2017core,zaruba2019ariane} over the accelerator cluster~\cite{rossi2017pulpcluster} to the on-chip communication fabric~\cite{kurth2021axi} have been taped out in multiple \glspl{asic}.

Programming models targeting heterogeneous computing are manifold, and we refer to \cite{mittal2015heterocompsurvey} for an overview.
In OpenCL~\cite{opencl2021c}, an application on the host submits separately-written kernels to be executed on an accelerator to a command queue.
OpenCL is imperative, meaning application programmers have to explicitly call functions to create buffers, transfer data, and start execution on an accelerator.
SYCL~\cite{sycl2021} extends OpenCL by enabling single-source heterogeneous programming and C++ AMP~\cite{cppamp2018} by relieving the programmers from explicit data transfers between host and device.
oneAPI Data Parallel C++ (DPC++)~\cite{oneapi2020} builds on SYCL to define functions that can be offloaded to devices, and an open-source LLVM implementation is in development.
OpenMP, supported natively by \hero{}, is declarative, meaning application programmers describe \emph{what} they want to do (e.g., offload a code section with data to an accelerator) while the compiler and runtime libraries take care of \emph{how} those actions happen.
OpenACC~\cite{openacc2020} goes even further: its directives describe the properties of a program (e.g., a parallel loop with independent iterations), and the toolchain and runtime libraries specialize the program to an accelerator.
In Clang, OpenACC is implemented by translation to OpenMP.
Through this, \hero{} also supports OpenACC.
\hero{}'s open-source LLVM-based toolchain will enable the community to construct complementary and alternative heterogeneous computing software stacks, while building on a solid open infrastructure.

Heterogeneous compilers have also been implemented by others.
Intel offers an OpenMP offloading compiler for its Xeon~Phi accelerators~\cite{intel2018offloading}, which differ from the host \gls{cpu} by accelerator-specific extensions.
Those extensions are only available through the proprietary Intel compiler, whereas \hero{}'s full toolchain is open source.
Research works on GCC~\cite{capotondi2018offloading} were the first to provide an open-source heterogeneous OpenMP toolchain, but GCC's offloading compilation is fundamentally limited to the same data model (e.g., 32-bit) for host and accelerators~\cite{kurth2020cc}.
Mixed-data-model heterogeneous compilation has been pioneered recently~\cite{kurth2020cc} with Clang/LLVM, and \hero{} integrates that work into its toolchain.

\section{Conclusion}%
\label{sec:conclusion}

\hero{} is a full-stack open-source\footnote{%
  \url{https://github.com/pulp-platform/hero}
} research platform for state-of-the-art heterogeneous computing:
\hero{} provides all hardware and software required to develop, compile, and run single-source, single-binary heterogeneous applications and seamlessly offload and share data from an application-class 64-bit host to a programmable 32-bit parallel accelerator.
Thus, \hero{} enables effective and accurate research from applications and algorithms down to microarchitecture.
Additionally, \hero{} comes with a novel \emph{AutoDMA} compiler plugin, which provides a solution to one of the most pressing problems of accelerators with software-managed memories:
without any code changes, AutoDMA tiles loops and infers \gls{dma} transfers, which leads to a speed-up of up to \SI{4.4}{\x} without any code changes and in most cases is only \SI{15}{\percent} slower than a handwritten implementation, which requires \SI{2.6}{\x} more code.

\hero{} enables research in various domains, and we know of ongoing projects that use \hero{} in high-performance computing, real-time processing, in-network processing, transprecision accelerators, and parallel programming.
We expect future work to evolve in the directions of larger scale-out accelerators, mixed and finer-grained coherency domains, and novel virtualization and communication technologies.
We are also working on a tape-out in a modern silicon technology.

\section*{Acknowledgments}
\small
The authors thank Koen Wolters for his work on the host software stack and the heterogeneous toolchain and Maxim Mattheeuws for porting \texttt{darknet} to \hero.
This work has been partially funded by
the ECSEL Joint Undertaking for the FRACTAL project under grant agreement no.~877056
and the Croatian-Swiss Research Programme (CSRP) for the Heterogeneous Computing Systems with Customized Accelerators (HCSCA) project under project no.~180625.

\bibliographystyle{IEEEtran}
\bibliography{IEEEabrv,main}

\begin{thebibliography}{10}
\providecommand{\url}[1]{#1}
\csname url@samestyle\endcsname
\providecommand{\newblock}{\relax}
\providecommand{\bibinfo}[2]{#2}
\providecommand{\BIBentrySTDinterwordspacing}{\spaceskip=0pt\relax}
\providecommand{\BIBentryALTinterwordstretchfactor}{4}
\providecommand{\BIBentryALTinterwordspacing}{\spaceskip=\fontdimen2\font plus
\BIBentryALTinterwordstretchfactor\fontdimen3\font minus
  \fontdimen4\font\relax}
\providecommand{\BIBforeignlanguage}[2]{{%
\expandafter\ifx\csname l@#1\endcsname\relax
\typeout{** WARNING: IEEEtran.bst: No hyphenation pattern has been}%
\typeout{** loaded for the language `#1'. Using the pattern for}%
\typeout{** the default language instead.}%
\else
\language=\csname l@#1\endcsname
\fi
#2}}
\providecommand{\BIBdecl}{\relax}
\BIBdecl

\bibitem{horowitz2014energyproblem}
M.~Horowitz, ``Computing's energy problem (and what we can do about it),'' in
  \emph{IEEE International Solid-State Circuits Conference (ISSCC)}, 2014, pp.
  10--14.

\bibitem{zahran2017heterogeneous}
\BIBentryALTinterwordspacing
M.~Zahran, ``Heterogeneous computing: Here to stay,'' \emph{Commun. ACM},
  vol.~60, no.~3, p. 42–45, Feb. 2017. [Online]. Available:
  \url{https://doi.org/10.1145/3024918}
\BIBentrySTDinterwordspacing

\bibitem{dally2020dsa}
\BIBentryALTinterwordspacing
W.~J. Dally \emph{et~al.}, ``Domain-specific hardware accelerators,''
  \emph{Commun. ACM}, vol.~63, no.~7, p. 48–57, Jun. 2020. [Online].
  Available: \url{https://doi.org/10.1145/3361682}
\BIBentrySTDinterwordspacing

\bibitem{nvidia2018xavier}
\BIBentryALTinterwordspacing
M.~Ditty \emph{et~al.}, ``{Nvidia's} {Xavier} {SoC},'' in \emph{IEEE/ACM
  HotChips 30}, 2018. [Online]. Available:
  \url{https://old.hotchips.org/hc30/1conf/1.12_Nvidia_XavierHotchips2018Final_814.pdf}
\BIBentrySTDinterwordspacing

\bibitem{tesla2019fsd}
\BIBentryALTinterwordspacing
P.~Bannon \emph{et~al.}, ``Compute and redundancy solution for the full
  self-driving computer,'' in \emph{IEEE/ACM HotChips 31}, 2019. [Online].
  Available:
  \url{https://old.hotchips.org/hc31/HC31_2.3_Tesla_Hotchips_ppt_Final_0817.pdf}
\BIBentrySTDinterwordspacing

\bibitem{amd2020renoir}
\BIBentryALTinterwordspacing
S.~Arora \emph{et~al.}, ``{AMD} {7nm} {Ryzen 4000} {APU} {Renoir},'' in
  \emph{IEEE/ACM HotChips 32}, 2020. [Online]. Available:
  \url{https://hotchips.org/assets/program/conference/day1/HotChips2020_Mobile_Processors_AMD_Renoir.pdf}
\BIBentrySTDinterwordspacing

\bibitem{hennessy2019goldenage}
\BIBentryALTinterwordspacing
J.~L. Hennessy \emph{et~al.}, ``A new golden age for computer architecture,''
  \emph{Commun. ACM}, vol.~62, no.~2, p. 48–60, Jan. 2019. [Online].
  Available: \url{https://doi.org/10.1145/3282307}
\BIBentrySTDinterwordspacing

\bibitem{reuther2020mlaccels}
A.~{Reuther} \emph{et~al.}, ``Survey of machine learning accelerators,'' in
  \emph{IEEE High Performance Extreme Computing Conference (HPEC)}, 2020, pp.
  1--12.

\bibitem{gui2019graphaccels}
C.-Y. Gui \emph{et~al.}, ``A survey on graph processing accelerators:
  Challenges and opportunities,'' \emph{Journal of Computer Science and
  Technology}, vol.~34, no.~2, pp. 339--371, 2019.

\bibitem{ubal2012multi2sim}
R.~{Ubal} \emph{et~al.}, ``{Multi2Sim}: A simulation framework for {CPU-GPU}
  computing,'' in \emph{21st IEEE International Conference on Parallel
  Architectures and Compilation Techniques (PACT)}, 2012, pp. 335--344.

\bibitem{power2015gem5gpu}
J.~{Power} \emph{et~al.}, ``{gem5-gpu}: A heterogeneous {CPU-GPU} simulator,''
  \emph{IEEE Computer Architecture Letters}, vol.~14, no.~1, pp. 34--36, 2015.

\bibitem{butko2016simulation}
A.~{Butko} \emph{et~al.}, ``Full-system simulation of {big.LITTLE} multicore
  architecture for performance and energy exploration,'' in \emph{2016 IEEE
  10th International Symposium on Embedded Multicore/Many-core Systems-on-Chip
  (MCSOC)}, 2016, pp. 201--208.

\bibitem{akram2019simulation}
A.~{Akram} \emph{et~al.}, ``A survey of computer architecture simulation
  techniques and tools,'' \emph{IEEE Access}, vol.~7, pp. 78\,120--78\,145,
  2019.

\bibitem{lee2016agile}
Y.~{Lee} \emph{et~al.}, ``An agile approach to building {RISC-V}
  microprocessors,'' \emph{IEEE Micro}, vol.~36, no.~2, pp. 8--20, 2016.

\bibitem{gray2016grviphalanx}
J.~{Gray}, ``{GRVI Phalanx}: A massively parallel {RISC-V} {FPGA} accelerator
  accelerator,'' in \emph{2016 IEEE 24th Annual International Symposium on
  Field-Programmable Custom Computing Machines (FCCM)}, 2016, pp. 17--20.

\bibitem{kamaleldin2020riscvfpga}
A.~{Kamaleldin} \emph{et~al.}, ``Towards a modular {RISC-V} based many-core
  architecture for {FPGA} accelerators,'' \emph{IEEE Access}, vol.~8, pp.
  148\,812--148\,826, 2020.

\bibitem{mantovani2020openesp}
P.~{Mantovani} \emph{et~al.}, ``Agile {SoC} development with {Open ESP},'' in
  \emph{2020 IEEE/ACM International Conference On Computer Aided Design
  (ICCAD)}, 2020, pp. 1--9.

\bibitem{balkind2020openpiton}
J.~{Balkind} \emph{et~al.}, ``{OpenPiton} at 5: A nexus for open and agile
  hardware design,'' \emph{IEEE Micro}, vol.~40, no.~4, pp. 22--31, 2020.

\bibitem{kurth2018herov1}
\BIBentryALTinterwordspacing
A.~Kurth \emph{et~al.}, ``{HERO}: An open-source research platform for {HW/SW}
  exploration of heterogeneous manycore systems,'' in \emph{Proceedings of the
  2nd Workshop on AutotuniNg and ADaptivity AppRoaches for Energy Efficient HPC
  Systems}, ser. ANDARE '18.\hskip 1em plus 0.5em minus 0.4em\relax New York,
  NY, USA: Association for Computing Machinery, 2018. [Online]. Available:
  \url{https://doi.org/10.1145/3295816.3295821}
\BIBentrySTDinterwordspacing

\bibitem{zaruba2019ariane}
F.~Zaruba \emph{et~al.}, ``The cost of application-class processing: Energy and
  performance analysis of a {Linux}-ready 1.7-{GHz} 64-bit {RISC-V} core in
  {22-nm} {FDSOI} technology,'' \emph{IEEE Transactions on Very Large Scale
  Integration (VLSI) Systems}, vol.~27, no.~11, pp. 2629--2640, 2019.

\bibitem{vogel2017onaccelptw}
\BIBentryALTinterwordspacing
P.~Vogel \emph{et~al.}, ``Efficient virtual memory sharing via on-accelerator
  page table walking in heterogeneous embedded {SoCs},'' \emph{ACM Trans.
  Embed. Comput. Syst.}, vol.~16, no.~5s, Sep. 2017. [Online]. Available:
  \url{https://doi.org/10.1145/3126560}
\BIBentrySTDinterwordspacing

\bibitem{mach2021fpnew}
S.~Mach \emph{et~al.}, ``{FPnew}: An open-source multiformat floating-point
  unit architecture for energy-proportional transprecision computing,''
  \emph{IEEE Transactions on Very Large Scale Integration (VLSI) Systems},
  vol.~29, no.~4, pp. 774--787, 2021.

\bibitem{conti2018xne}
F.~{Conti} \emph{et~al.}, ``{XNOR} neural engine: A hardware accelerator {IP}
  for 21.6-{fJ}/op binary neural network inference,'' \emph{IEEE Transactions
  on Computer-Aided Design of Integrated Circuits and Systems}, 2018.

\bibitem{kurth2020cc}
\BIBentryALTinterwordspacing
A.~Kurth \emph{et~al.}, ``Mixed-data-model heterogeneous compilation and
  {OpenMP} offloading,'' in \emph{Proceedings of the 29th International
  Conference on Compiler Construction}, ser. CC 2020.\hskip 1em plus 0.5em
  minus 0.4em\relax New York, NY, USA: Association for Computing Machinery,
  2020, p. 119–131. [Online]. Available:
  \url{https://doi.org/10.1145/3377555.3377891}
\BIBentrySTDinterwordspacing

\bibitem{kurth2018iccd}
------, ``Scalable and efficient virtual memory sharing in heterogeneous {SoCs}
  with {TLB} prefetching and {MMU}-aware {DMA} engine,'' in \emph{2018 IEEE
  36th International Conference on Computer Design (ICCD)}, 2018, pp. 292--300.

\bibitem{kurth2021axi}
------, ``An open-source platform for high-performance non-coherent on-chip
  communication,'' \emph{IEEE Transactions on Computers}, 2021.

\bibitem{cavalcante2020bridge}
\BIBentryALTinterwordspacing
M.~Cavalcante \emph{et~al.}, ``Design of an open-source bridge between
  non-coherent burst-based and coherent cache-line-based memory systems,'' in
  \emph{Proceedings of the 17th ACM International Conference on Computing
  Frontiers}, ser. CF '20.\hskip 1em plus 0.5em minus 0.4em\relax New York, NY,
  USA: Association for Computing Machinery, 2020, p. 81–88. [Online].
  Available: \url{https://doi.org/10.1145/3387902.3392631}
\BIBentrySTDinterwordspacing

\bibitem{forsberg2020heprem}
B.~Forsberg \emph{et~al.}, ``Heprem: A predictable execution model for
  gpu-based heterogeneous socs,'' \emph{IEEE Transactions on Computers},
  vol.~70, no.~1, pp. 17--29, 2020.

\bibitem{openmp45}
\emph{{OpenMP} Application Programming Interface, Version 4.5}, {OpenMP}
  Architecture Review Board, 2015.

\bibitem{hsa2016}
W.-M.~W. Hwu, Ed., \emph{Heterogeneous System Architecture}.\hskip 1em plus
  0.5em minus 0.4em\relax Morgan Kaufmann Publishers, 2016.

\bibitem{posix2017}
\emph{{POSIX.1-2017} ({IEEE Std 1003.1-2017})}, {IEEE} and {The Open Group},
  2018.

\bibitem{herter2014allocation}
J.~Herter, ``Timing-predictable memory allocation in hard real-time systems,''
  Ph.D. dissertation, Universit\"{a}t des Saarlandes, 2014.

\bibitem{kirienko2020o1heap}
\BIBentryALTinterwordspacing
P.~Kirienko, ``Constant-complexity deterministic memory allocator (heap) for
  hard real-time high-integrity embedded systems,'' 2020. [Online]. Available:
  \url{https://github.com/pavel-kirienko/o1heap}
\BIBentrySTDinterwordspacing

\bibitem{gautschi2017core}
M.~Gautschi \emph{et~al.}, ``Near-threshold {RISC-V} core with {DSP} extensions
  for scalable {IoT} endpoint devices,'' \emph{IEEE Transactions on Very Large
  Scale Integration (VLSI) Systems}, vol.~25, no.~10, pp. 2700--2713, 2017.

\bibitem{zaruba2020snitch}
F.~Zaruba \emph{et~al.}, ``{Snitch}: A tiny pseudo dual-issue processor for
  area and energy efficient execution of floating-point intensive workloads,''
  \emph{IEEE Transactions on Computers}, pp. 1--1, 2020.

\bibitem{grauer2012auto}
S.~Grauer-Gray \emph{et~al.}, ``{Auto-tuning a high-level language targeted to
  GPU codes},'' in \emph{2012 Innovative Parallel Computing (InPar)}.\hskip 1em
  plus 0.5em minus 0.4em\relax IEEE, 2012, pp. 1--10.

\bibitem{redmon2016yolo}
J.~Redmon \emph{et~al.}, ``{You Only Look Once}: Unified, real-time object
  detection,'' in \emph{2016 IEEE Conference on Computer Vision and Pattern
  Recognition (CVPR)}, 2016, pp. 779--788.

\bibitem{fleming1986geomean}
\BIBentryALTinterwordspacing
P.~J. Fleming \emph{et~al.}, ``How not to lie with statistics: The correct way
  to summarize benchmark results,'' \emph{Commun. ACM}, vol.~29, no.~3, p.
  218–221, Mar. 1986. [Online]. Available:
  \url{https://doi.org/10.1145/5666.5673}
\BIBentrySTDinterwordspacing

\bibitem{littlefair2001code}
T.~Littlefair, ``An investigation into the use of software code metrics in the
  industrial software development environment,'' Ph.D. dissertation, Edith
  Cowan University, 2001.

\bibitem{grosser2012polly}
T.~Grosser \emph{et~al.}, ``Polly—performing polyhedral optimizations on a
  low-level intermediate representation,'' \emph{Parallel Processing Letters},
  vol.~22, no.~04, p. 1250010, 2012.

\bibitem{cadence2021palladium}
\emph{{Palladium Z2 Enterprise Emulation Platform} -- Enabling design teams to
  debug and accelerate verification of complex {IPs} to multi-billion-gate
  {SoC} designs}, Cadence, 2021.

\bibitem{siemens2021veloce}
``Emulation -- a job management strategy to maximize use,'' Siemens, White
  Paper, 2021.

\bibitem{synopsys2021zebu}
\emph{{ZeBu} {EP1}: Industry's Fastest Billion Gates Emulation System},
  Synopsys, 2021.

\bibitem{balkind2016openpiton}
\BIBentryALTinterwordspacing
J.~Balkind \emph{et~al.}, ``{OpenPiton}: An open source manycore research
  framework,'' \emph{SIGPLAN Not.}, vol.~51, no.~4, p. 217–232, Mar. 2016.
  [Online]. Available: \url{https://doi.org/10.1145/2954679.2872414}
\BIBentrySTDinterwordspacing

\bibitem{balkind2020byoc}
\BIBentryALTinterwordspacing
------, ``{BYOC}: A "bring your own core" framework for heterogeneous-{ISA}
  research,'' in \emph{Proceedings of the Twenty-Fifth International Conference
  on Architectural Support for Programming Languages and Operating Systems},
  ser. ASPLOS '20.\hskip 1em plus 0.5em minus 0.4em\relax New York, NY, USA:
  Association for Computing Machinery, 2020, p. 699–714. [Online]. Available:
  \url{https://doi.org/10.1145/3373376.3378479}
\BIBentrySTDinterwordspacing

\bibitem{zhang2020meg}
\BIBentryALTinterwordspacing
J.~Zhang \emph{et~al.}, ``{MEG}: A {RISCV}-based system emulation
  infrastructure for near-data processing using {FPGAs} and high-bandwidth
  memory,'' \emph{ACM Trans. Reconfigurable Technol. Syst.}, vol.~13, no.~4,
  Sep. 2020. [Online]. Available: \url{https://doi.org/10.1145/3409114}
\BIBentrySTDinterwordspacing

\bibitem{wang2014dart}
D.~Wang \emph{et~al.}, ``{DART}: A programmable architecture for {NoC}
  simulation on {FPGAs},'' \emph{IEEE Transactions on Computers}, vol.~63,
  no.~3, pp. 664--678, 2014.

\bibitem{prasad2021fpganoc}
B.~P. Prasad \emph{et~al.}, ``{FPGA} friendly {NoC} simulation acceleration
  framework employing the hard blocks,'' \emph{Computing}, pp. 1--23, 2021.

\bibitem{kamali2018ducnoc}
H.~M. Kamali \emph{et~al.}, ``{DuCNoC}: A high-throughput {FPGA}-based {NoC}
  simulator using dual-clock lightweight router micro-architecture,''
  \emph{IEEE Transactions on Computers}, vol.~67, no.~2, pp. 208--221, 2018.

\bibitem{karandikar2018firesim}
S.~Karandikar \emph{et~al.}, ``{FireSim}: {FPGA}-accelerated cycle-exact
  scale-out system simulation in the public cloud,'' in \emph{2018 ACM/IEEE
  45th Annual International Symposium on Computer Architecture (ISCA)}, 2018,
  pp. 29--42.

\bibitem{karandikar2020fireperf}
\BIBentryALTinterwordspacing
------, ``{FirePerf}: {FPGA}-accelerated full-system hardware/software
  performance profiling and co-design,'' in \emph{Proceedings of the
  Twenty-Fifth International Conference on Architectural Support for
  Programming Languages and Operating Systems}, ser. ASPLOS '20.\hskip 1em plus
  0.5em minus 0.4em\relax New York, NY, USA: Association for Computing
  Machinery, 2020, p. 715–731. [Online]. Available:
  \url{https://doi.org/10.1145/3373376.3378455}
\BIBentrySTDinterwordspacing

\bibitem{huang2019centrifuge}
Q.~Huang \emph{et~al.}, ``{Centrifuge}: Evaluating full-system {HLS}-generated
  heterogenous-accelerator {SoCs} using {FPGA}-acceleration,'' in \emph{2019
  IEEE/ACM International Conference on Computer-Aided Design (ICCAD)}, 2019,
  pp. 1--8.

\bibitem{giri2021openesp}
D.~Giri \emph{et~al.}, ``Accelerator integration for open-source {SoC}
  design,'' \emph{IEEE Micro}, vol.~41, no.~4, pp. 8--14, 2021.

\bibitem{kurth2017herov1}
A.~Kurth \emph{et~al.}, ``{HERO:} heterogeneous embedded research platform for
  exploring {RISC-V} manycore accelerators on {FPGA},'' in \emph{Proceedings of
  the First International Workshop on Computer Architecture Research with
  RISC-V}, ser. CARRV '17, 2017.

\bibitem{zhangxi2010fame}
\BIBentryALTinterwordspacing
Z.~Tan \emph{et~al.}, ``A case for {FAME}: {FPGA} architecture model
  execution,'' \emph{SIGARCH Comput. Archit. News}, vol.~38, no.~3, p.
  290–301, Jun. 2010. [Online]. Available:
  \url{https://doi.org/10.1145/1816038.1815999}
\BIBentrySTDinterwordspacing

\bibitem{zuckerman2021cohmeleon}
\BIBentryALTinterwordspacing
J.~Zuckerman \emph{et~al.}, ``{Cohmeleon}: Learning-based orchestration of
  accelerator coherence in heterogeneous {SoCs},'' in \emph{MICRO-54: 54th
  Annual IEEE/ACM International Symposium on Microarchitecture}, ser. MICRO
  '21.\hskip 1em plus 0.5em minus 0.4em\relax New York, NY, USA: Association
  for Computing Machinery, 2021, p. 350–365. [Online]. Available:
  \url{https://doi.org/10.1145/3466752.3480065}
\BIBentrySTDinterwordspacing

\bibitem{angepat2014fpgaemul}
\BIBentryALTinterwordspacing
H.~Angepat \emph{et~al.}, ``{FPGA}-accelerated simulation of computer
  systems,'' \emph{Synthesis Lectures on Computer Architecture}, vol.~9, no.~2,
  pp. 1--80, 2014. [Online]. Available:
  \url{https://doi.org/10.2200/S00586ED1V01Y201407CAC029}
\BIBentrySTDinterwordspacing

\bibitem{amid2020chipyard}
A.~Amid \emph{et~al.}, ``{Chipyard}: Integrated design, simulation, and
  implementation framework for custom {SoCs},'' \emph{IEEE Micro}, vol.~40,
  no.~4, pp. 10--21, 2020.

\bibitem{gray20192grviphalanx}
J.~{Gray}, ``{2GRVI Phalanx}: Towards kilocore {RISC-V} {FPGA} accelerators
  with {HBM2} {DRAM},'' in \emph{IEEE/ACM HotChips 30}, 2019.

\bibitem{abdelhamid2021dragon}
R.~B. Abdelhamid \emph{et~al.}, ``A highly-efficient and tightly-connected
  many-core overlay architecture,'' \emph{IEEE Access}, vol.~9, pp.
  65\,277--65\,292, 2021.

\bibitem{rossi2017pulpcluster}
D.~Rossi \emph{et~al.}, ``Energy-efficient near-threshold parallel computing:
  The {PULPv2} cluster,'' \emph{IEEE Micro}, vol.~37, no.~5, pp. 20--31, 2017.

\bibitem{mittal2015heterocompsurvey}
\BIBentryALTinterwordspacing
S.~Mittal \emph{et~al.}, ``A survey of {CPU-GPU} heterogeneous computing
  techniques,'' \emph{ACM Comput. Surv.}, vol.~47, no.~4, Jul. 2015. [Online].
  Available: \url{https://doi.org/10.1145/2788396}
\BIBentrySTDinterwordspacing

\bibitem{opencl2021c}
\emph{The {OpenCL} {C} Specification v3.0.7}, {Khronos} {OpenCL} Working Group,
  Apr. 2021.

\bibitem{sycl2021}
\emph{{SYCL} 2020 Specification (revision 3)}, {Khronos} {SYCL} Working Group,
  Mar. 2021.

\bibitem{cppamp2018}
\emph{{C++} {Accelerated Massive Parallelism} {(C++ AMP)}}, Microsoft Corp.,
  Nov. 2018.

\bibitem{oneapi2020}
\emph{{oneAPI} Specification, Release {1.0-rev-3}}, Intel Corp., Nov. 2020.

\bibitem{openacc2020}
\emph{The {OpenACC} Application Programming Interface v3.1}, The {OpenACC}
  Organization, Nov. 2020.

\bibitem{intel2018offloading}
``Migrating offloading software to {Intel} {Xeon Phi} processor,'' Intel Corp.,
  White Paper, Feb. 2018.

\bibitem{capotondi2018offloading}
A.~Capotondi \emph{et~al.}, ``Runtime support for multiple offload-based
  programming models on clustered manycore accelerators,'' \emph{IEEE
  Transactions on Emerging Topics in Computing}, vol.~6, no.~3, pp. 330--342,
  2018.

\end{thebibliography}

\newenvironment{biography}[2][example-image-a]{%
  \begin{IEEEbiography}[{%
    \includegraphics[height=.83in,width=.6457in]{#1}%
  }]{#2}%
}{%
  \end{IEEEbiography}%
}
\newcommand{\lucaphd}{He is currently pursuing a PhD degree in the Digital Circuits and Systems group of Prof.\ Benini.}
\newcommand{\ethgrad}[2]{received his BSc and MSc degree in electrical engineering and information technology from ETH Zurich in #1 and #2, respectively.}
\newcommand{\researchinterests}[1]{His research interests include #1.}

\begin{biography}[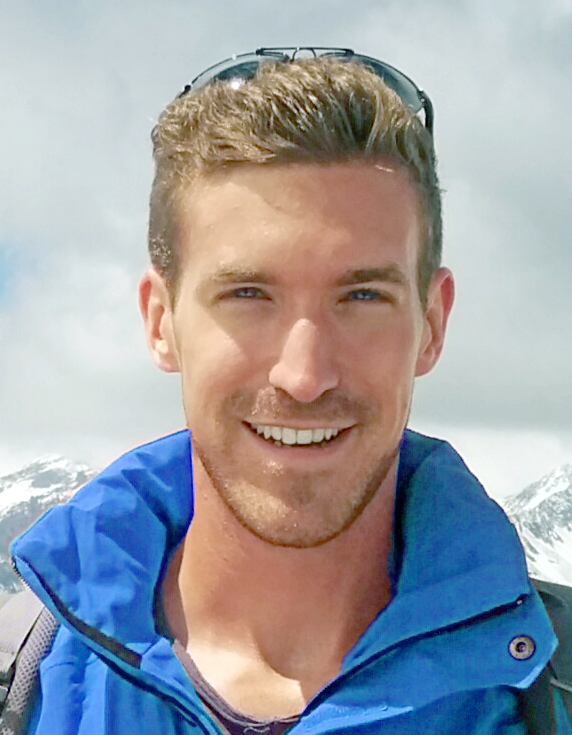]{Andreas Kurth}
    \ethgrad{2014}{2017}
    \lucaphd{}
    \researchinterests{%
        the architecture and programming of heterogeneous \acrshortpl{soc} and %
        accelerator-rich computing systems}
\end{biography}

\begin{biography}{Bj\"{o}rn Forsberg} received his PhD from ETH Zurich in 2021, working on compiler and software-centric techniques for real-time execution on embedded heterogeneous systems.
His interests lie at new and enabling technology at the hardware/software boundary, and its impact on programmability, real-time, and performance.
He received his MSc degree from Uppsala Universitet in 2015.
\end{biography}

\begin{biography}[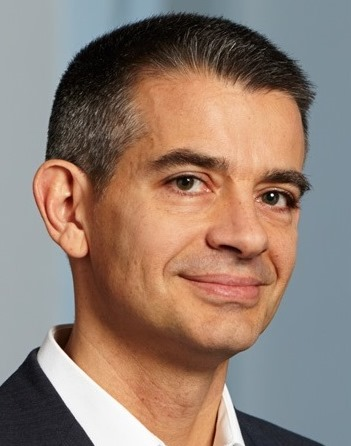]{Luca Benini}
    (F'07) holds the chair of Digital Circuits and Systems at ETH Zurich and is Full Professor at the Università di Bologna.
    Dr.\ Benini’s research interests are in energy-efficient computing systems design, from embedded to high-performance.
    He has published more than 1000 peer-reviewed papers and five books.
    He is a Fellow of the ACM and a member of Academia Europaea.
\end{biography}

\end{document}